\documentclass[12pt]{article}
 \pdfoutput=1

\usepackage{natbib}

\usepackage{tikz}
\usetikzlibrary{positioning}

\usepackage{amsmath,amssymb, amsfonts, tabularx, caption, bbm, lmodern}
\usepackage{soul}
\newcommand*{\rom}[1]{\expandafter\@slowromancap\romannumeral #1@}
\newcommand{\xmark}{\ding{55}}
\newcommand{\cmark}{\ding{51}}
\usepackage{relsize}

\usepackage{graphicx}
\usepackage{float}

\usepackage{multirow}
\usepackage{makecell}
\usepackage{verbatimbox}
\newcommand{\indep}{\rotatebox[origin=c]{90}{$\models$}}
\newcolumntype{P}[1]{>{\centering\arraybackslash}p{#1}}
\usepackage{rotating}

\usepackage{pifont}
{
\newtheorem{assumption}{Assumption}
\newtheorem{proposition}{Proposition}

}

\usepackage{DNstyle}
\newcommand{\Smono}{S-Monotonicity }
\newcommand{\SmonoNO}{S-Monotonicity}
\newcommand{\Dmono}{D-Monotonicity }
\newcommand{\DmonoNO}{D-Monotonicity}
\def \sface {SF\text{-}ACE}

\usepackage{url}
\newcommand{\blind}{1}

\begin{document}


\def\spacingset#1{\renewcommand{\baselinestretch}%
{#1}\small\normalsize} \spacingset{1}

\if1\blind
	{
\title{\bf The Subtype-Free Average Causal Effect for Heterogeneous Disease Etiology}
\author{Amit Sasson$^1$,
		Molin Wang$^{2,3,4}$
		Shuji Ogino$^{3,5,6}$
		and 
		Daniel Nevo$^{1}$\hspace{-.2cm} \thanks{ \noindent  danielnevo@gmail.com \hfill \break
1 Department of Statistics and Operations Research, Tel Aviv University, Tel Aviv, Israel. \hfill \break
2 Department of Biostatistics, Harvard T.H. Chan School of Public Health, Boston, MA, USA
\hfill \break
3 Department of Epidemiology, Harvard T.H. Chan School of Public Health, Boston, MA, USA
\hfill \break
4 Channing Division of Network Medicine, Department of Medicine, Brigham and Women's Hospital and Harvard Medical School, Boston, MA, USA
\hfill \break
5 Broad Institute of MIT and Harvard, Cambridge, MA, USA
\hfill \break
6 Program in MPE Molecular Pathological Epidemiology, Department of Pathology, Brigham and Women’s Hospital Harvard Medical School, Boston, MA, USA
\hfill \break
This research was supported by the Israel Science Foundation (ISF grant No. 827/21); U.S. National Institutes of Health (NIH) grants R35 CA197735, R01 CA151993, UM1 CA186107, P01 CA87969, P01 CA55075, UM1 CA167552,  U01 CA16755; and Cancer Research UK Grand Challenge Award (UK C10674/A27140)}}
		\date{}
		\maketitle
} \fi
	
\if0\blind
{
	\bigskip
	\bigskip
	\bigskip
	\begin{center}
	{\LARGE\bf The Subtype-Free Average Causal Effect for Heterogeneous Disease Etiology}
	\end{center}
	\medskip
} \fi

\bigskip	
	
\begin{abstract} 
Studies have shown that the effect an exposure may have on a disease can vary for different subtypes of the same disease. However, existing approaches to estimate and compare these effects largely overlook causality. 
In this paper, we study the effect smoking may have on having colorectal cancer subtypes defined by a trait known as microsatellite instability (MSI). We use principal stratification to propose an alternative causal estimand, the Subtype-Free Average Causal Effect (SF-ACE). The SF-ACE is the causal effect of the exposure among those who would be free from other disease subtypes under any exposure level. We study non-parametric identification of the SF-ACE, and discuss different monotonicity assumptions, which are more nuanced than in the standard setting. As is often the case with principal stratum effects, the assumptions underlying the identification of the SF-ACE from the data are untestable and can be too strong. Therefore, we also develop sensitivity analysis methods that relax these assumptions. We present three different estimators, including a doubly-robust estimator, for the SF-ACE. We implement our methodology for data from two large cohorts to study the heterogeneity in the causal effect of smoking on colorectal cancer with respect to MSI subtypes.
\end{abstract}

\noindent%
\vfill	

\newpage
\spacingset{1.9} 

\section{Introduction} 
\label{section:Introduction}


In recent decades, it has become clear that many diseases that share general clinical characteristics evolve through a range of heterogeneous molecular pathologic processes, which may be affected differently by the same exposure.
For this reason, classifying a disease into subtypes according to criteria based on molecular characteristics can improve our etiologic understanding of the disease \citep{ogino2016role}.
	
In this paper, we consider the causal effect  smoking may have on having colorectal cancer (CRC). 
One well-established CRC subtype classification  is based on  microsatellite instability (MSI). The two  mutually-exclusive subtypes are MSI-high (approximately 15\% of the cases) and non-MSI-high (approximately 85\%). 
Smoking has been found to be a strong risk factor for the MSI-high subtype, while  weaker evidence was found that smoking is a risk factor for the non-MSI-high subtype \citep{carr2018lifestyle,Amitay2020smoking}.

To study etiologic heterogeneity, researchers often use a multinomial regression model in which being healthy or having each of the disease subtypes 
form the possible values of the outcome. However,  the multinomial regression parameters do not correspond to well-defined causal effects, because the multinomial regression parameters are equivalent to the parameters obtained from a series of logistics regressions, each comparing one disease subtype to the healthy controls, resulting in selection bias \citep{nevo2021reflection}.  

This form of bias is not limited to subtype comparisons and is not unique to multinomial regression. More generally, competing events create a challenge in making causal statements. Section \ref{SubSec:ReviewCausal}  reviews related approaches and considers their applicability in our setup \citep{young2020causal,stensrud2020separable,nevo2021reflection}. 
	
In this paper, we use data from  two large cohorts  to study well-defined and relevant estimands representing the effect of smoking on the two CRC subtypes. We propose an alternative estimand, inspired by the Survivor Average Causal Effect ($SACE$) \citep{zhang2003estimation,rubin2006causal,zehavi2021matching}. The $SACE$ is a causal effect typically-used to overcome the problem of truncation by death, which occurs when an outcome of interest is undefined for individuals who died before their outcome could be measured.  The $SACE$ is defined using principal stratification \citep{robins1986new,frangakis2002principal} as the average causal effect of a treatment among individuals that would have survived under both treatment/exposure values.  The $SACE$ is not identifiable from the data under standard assumptions. Approaches to identify and estimate $SACE$ include relying on additional strong assumptions, performing sensitivity analyses, and obtaining bounds. 

Our proposed approach is based on the same idea as the $SACE$. First, we define principal strata with respect to the potential outcomes of other disease subtypes. Then, we define the Subtype-Free Average Causal Effect ($\sface$) to be the average causal effect of the exposure on one disease subtype among the individuals who would have been free of the other MSI disease subtype under either exposure level (smoking status). 

One commonly-made assumption in truncation-by-death problems is \textit{Monotonicity}, namely that treatment/exposure cannot hurt survival \citep{zhang2003estimation}. In the setting of disease heterogeneity with two subtypes, the definition of Monotonicity is more nuanced for two reasons. First, the assumption needs to be made (or not) for each subtype separately. Second, the disease  could occur under both exposure statuses, but the subtype may differ. We term this phenomenon exposure-induced subtype switching and show it plays a key role in the (lack of) identification of the $\sface$. We further show that unlike the $SACE$ setting, the $\sface$ is non-parametrically identifiable under certain Monotonicity assumptions.  Under weaker Monotonicity assumptions, the $\sface$ is no longer identifiable. For this scenario, we develop a sensitivity analysis  approach as a function of  the unidentifiable exposure-induced subtype switching probabilities.

We consider three  $\sface$ estimators. The first  two, standardization-based estimator and an inverse probability of treatment weighting (IPTW) estimator, may rely on parametric modeling assumptions. We therefore present a doubly-robust (DR) estimator that will be consistent if at least one of two models, but not necessarily both,  is correctly specified. 

The rest of the paper is organized as follows. Section \ref{Sec:DataDesc} describes the epidemiologic problem and the available data. Section \ref{Sec:Prelim} presents the notations and the causal framework. Section \ref{Sec:ReviewExisting} reviews existing approaches for etiologic disease heterogeneity and discuss available causal estimands. Section \ref{Sec:SFACE} presents a detailed study of the proposed causal estimand, the $\sface$, including its interpretation, assumptions needed for its identification, sensitivity analyses and estimation. Section \ref{Sec:Sims} summarizes the results of our simulation studies. Section \ref{Sec:DataAnalysis} presents various analyses of the data to study the causal effect of smoking on the two CRC subtypes. Final conclusions are given in Section \ref{Sec:Discuss}.  The \textbf{R} package \texttt{TheSFACE} is available from CRAN implements our methodology. Reproducibility materials are available from https://github.com/amitSasson/SFACE\_Reproduce.
	
\section{Data and problem description}
\label{Sec:DataDesc}

We use data from two large US cohorts. The  Nurses Health Study (NHS) was established in 1976 and consists of female nurses aged 30--55 at the beginning of follow-up. The Health Professionals Follow-up Study (HPFS) was established in 1986 and consists of male health professionals aged 40--75 at the beginning of follow-up. Participants answered biennial questionnaires about lifestyle, medical and other health-related information every two years. 
CRC diagnoses were reported by study participants.  Formalin-fixed paraffin-embedded tumor tissue specimens were retrieved from hospitals across the US, and a CRC diagnosis was confirmed by the study pathologist (S.O.), who also conducted analyses to determine the MSI status. Further details can be found in \cite{ugai2022smoking} and references therein.

We set the baseline to be age 60, and define the exposure to be ever versus never smoking by age 60. Such binary exposure definitions are often used to minimize recall bias and measurement error. The outcome was CRC by age 70, subtyped by MSI status. Age 70 was chosen to minimize the impact of death as a competing event.

We removed from the data people who were older than 60 when enrolled in the study, or people who died, had CRC, or were lost to follow up before age 60. As in \cite{ugai2022smoking}, we also removed people with missing baseline smoking data.  In total, we were left with 114,947 people at baseline.  Then, we removed 5,132 people (4.4\%) who have died, without being diagnosed, prior to age 70. We were left with a final sample of 109,815. In this sample, 17,854 people were CRC-free and younger than 70 when the last questionnaire was collected (out of which 94.8\% were older than 65). We treated them as CRC-free at age 70, and examined how it may affected our results by considering alternative analyses.

In our sample, the number of exposed (ever smoking) was 58,432 (53.3\%) and the number of CRC cases by age 70 was  961 (0.9\%). Out of them, 358 were diagnosed with non-MSI-high CRC (37.3\%) and 61 were diagnosed with MSI-high CRC (6.3\%). The subtype was missing for the remaining 542 (56.4\%) diagnosed participants, as is often the case in such studies \citep{nevo2018competing,liu2018utility}. To have a clear presentation of the issues at end, we develop our theory and methodology under the assumption that subtypes are known for all disease cases. We address the issue of missing subtypes in our analysis in Section \ref{Sec:DataAnalysis} using inverse probability weighting \citep{liu2018utility}.

 The NHS and HPFS cohorts include a vast number of covariates collected through biennial questionnaires. Table C.7 in the Appendix provides information on key baseline confounders we adjusted for in our analyses:  gender (male/female), regular aspirin use (yes/no), first-degree family history of CRC (yes/no), history of lower endoscopy (yes/no), physical activity level (mean metabolic equivalent task score hours per week), body mass index (categorical), alcohol intake (g/day) and total calorie intake (kcal/day). 
\section{Preliminaries}
\label{Sec:Prelim}

Using the potential outcomes framework, let $Y_i(a)$ denote the potential disease status had the exposure of individual $i=1,...,n$ been set to $A_i=a$ for $a=0,1$. In our study, $A_i=0$ indicates no smoking by age 60 (as captured by the cohort data), while $A_i=1$ indicates ever smoking by age 60.  If individual $i$ would have  been diagnosed with disease subtype $k=1,...,K$ under exposure value $A_i=a$, then $Y_i(a) = k$, and if they would be disease-free under $A_i=a$, then $Y_i(a) = 0$. Let also $Y_i^{(k)}(a) = \bbmI\{Y_i(a) = k\}$ indicate whether individual $i$ is diagnosed with disease type $k$ under $A_i=a$. Note that because the subtypes are mutually exclusive, $\sum_{k=1}^{K}Y_i^{(k)}(a)\le 1$. In our study, $K=2$, subtype 1 is non-MSI-high CRC, and subtype 2 is MSI-high CRC. So, for example, $Y_i^{(1)}(0)=1$ if person $i$ would have been diagnosed with non-MSI-high CRC by age 70 had they not been smoking by age 60.

Throughout the paper, we assume there is no interference between individuals and that there are no multiple versions of the exposure leading to different potential outcomes. Under these assumptions, known collectively as the stable unit treatment value assumption (\textit{SUTVA}), the observed disease status for person $i$, hereafter denoted by $Y_i$, with observed exposure value $A_i$, equals to the appropriate potential outcome for that person. That is,  $Y_i=Y_i(A_i)$ (and $Y^{(k)}_i=Y^{(k)}_i(A_i)$). The latter is also known as the \textit{consistency} assumption.

As in many observational studies, we also assume \textit{weak ignorability}, namely that rich enough data were collected as part of the NHS and HPFS cohorts such that for a vector of measured covariates $\bX$, we have $A \indep Y(a)\ |\ \bX$ for $a=0,1$. Another typically needed assumption for identification of causal effects is \textit{positivity}, meaning that the exposure assignment probability $e(\bx)=\Pr(A=a|\bX=\bx)$ is positive for $a=0,1$ and for all $\bx$. 
	
In summary, the observed data consists of $n$ i.i.d copies sampled from the distribution of $(\bX, A, Y^{(1)},...,Y^{(K)})$, that can be equivalently represented as $(\bX, A, Y)$.



\section{Review of existing approaches}	
\label{Sec:ReviewExisting}

A principled approach to causal inference typically starts with a definition of causal effects of interest, followed by the assumptions needed for their identification from observed data, and only then statistical models are discussed. However, as the existing approaches employed in practice for  studying etiologic heterogeneity are model-based and are not grounded in the causal inference paradigm, we start with reviewing these models (Section \ref{SubSec:ReviewStat}). Then, we discuss relevant causal effects arising in the presence of competing events (Section \ref{SubSec:ReviewCausal}).


\subsection{Statistical models for disease heterogeneity}
\label{SubSec:ReviewStat}
Let $\pi_k(a,\bx)=\Pr(Y = k|A = a, \bX =  \bx)$, $k=0,...,K$, be the probability of subtype $k$ disease (or being subtype-free for $k=0$) among those with exposure $A=a$ and confounders $\bX=\bx$. Let also $\bpi(a,\bx)=(\pi_0(a,\bx),...,\pi_K(a,\bx))$.
A commonly-used approach for studying etiologic heterogeneity is built upon the multinomial regression model 
\begin{equation}
\label{Eq:MultinoModel}
\pi_k(a,\bx) = \frac{\exp(\alpha_k + \beta_k a + \bgamma_k^T \bx)}{1 + \sum_{j=1}^{K} \exp(\alpha_j + \beta_j a + \bgamma_j^T  \bx)},
\end{equation}
for $k=1,...,K$ and $\pi_0(a,\bx) = 1 - \sum_{k=1}^{K} \pi_k(a,\bx)$. Under this model, $\exp(\beta_k)$ is the odds ratio (OR) between the exposure $A$ and the observed subtype-specific outcome $Y^{(k)}$ compared with not having the disease, that is,
$
\exp(\beta_k) = \frac{\pi_k(1,\bx)}{\pi_0(1,\bx)}\times \frac{\pi_0(0,\bx)}{\pi_k(0,\bx)}.
$
The parameter $\exp(\beta_k)$ also approximates the risk ratio (RR) when the disease is rare, as in our study.
Model \eqref{Eq:MultinoModel} was also extended for studies within which subtypes were not a-priori known and may be defined using multiple, possibly mis-meausred, biomarkers \citep{chatterjee2004two,wang2015meta,nevo2016accounting}. See also \cite{wang2016statistical,zabor2017comparison} for a survey of and comparison between methods based on this and similar models.

Estimation of Model \eqref{Eq:MultinoModel} parameters is typically carried out by standard maximum likelihood approach, with some exceptions when further complications arise \citep[e.g.,][]{chatterjee2004two,nevo2016accounting}. In practice, researchers are often interested in whether there is heterogeneity in the effects of $A$ across disease subtypes. When the null hypothesis $H_0: \beta_1 = ... =  \beta_{K}$ is rejected, researchers conclude that the exposure $A$ contributes to the risk of at least one subtype differently from its contribution to the risk of other subtypes.

A recently-proposed alternative approach postulates $K$ logistic regressions, each comparing a subtype $k$ disease to the alternative of being free of that subtype,  that is,  being either disease-free or diagnosed with the disease, but of a different subtype \citep{sun2017multinomial}. Formally, these models can be expressed by
\begin{equation}
\label{Eq:SunModel}
\pi_k(a,\bx) = \frac{\exp(\widetilde{\alpha}_k + \widetilde{\beta}_k a + \widetilde{\bgamma}_k^T \bx)}{1 + \exp(\widetilde{\alpha}_k + \widetilde{\beta}_k a + \widetilde{\bgamma}_k^T  \bx)},
\end{equation}
for $k=1,...,K$. Under this model, $\exp(\widetilde{\beta}_k)$ is the OR between the exposure $A$ and the observed indicator of subtype-specific outcome $Y^{(k)}$; that is,
$
\exp(\widetilde{\beta}_k) = \frac{\pi_k(1,\bx)}{1-\pi_k(1,\bx)}\times \frac{1-\pi_k(0,\bx)}{\pi_k(0,\bx)}.
$
Because fitting these models separately may result in $\sum_{k=0}^{K}\pi_k\ne 1$, estimation of the parameters in \eqref{Eq:SunModel} calls for constrained maximization of the models' likelihood. A more computationally attractive approach, developed by   \cite{sun2017multinomial}, uses constrained Bayesian estimation, and limits the posterior distribution of the parameter vector to have positive mass only for parameter values such that $\sum_{k=0}^{K}\pi_k = 1$.

Comparing models \eqref{Eq:MultinoModel} and \eqref{Eq:SunModel}, it has been recently argued \citep{sun2017multinomial} that variation in $\beta_k$ across the subtypes (across $k$) is not a satisfactory measure of etiologic heterogeneity, because the parameter space of a multinomial regression model requires a risk factor for one subtype to be a risk factor for the other subtypes, and that $\widetilde{\beta}_k$ may be preferred, although this issue was also debated \citep{begg2018re, sun2018reply}.


\subsection{Causal effects for competing risks}
\label{SubSec:ReviewCausal}

To fix ideas, we focus on the effect of $A$ on $Y^{(1)}$, the first subtype, and start with a na\"ive approach that does not correspond to a causal effect. Researchers may opt for a comparison of the subtype 1 proportion of cases at each exposure level only among individuals who are free of other disease subtypes. When researchers also adjust for the confounders $\bX$, their statistical estimand on the difference scale is
\begin{equation}
\label{Eq:NaiveCond}
\bbE_{\bX}\{\bbE[Y^{(1)}| A = 1, \bX,  Y^{(2)} = ... = Y^{(K)} = 0] -\bbE[Y^{(1)}| A = 0, \bX, Y^{(2)} = ... = Y^{(K)} = 0]\},
\end{equation}
and can be analogously defined on the RR or OR scales. Because model \eqref{Eq:MultinoModel} is equivalent to $K$ separate logistic regression models, each comparing one disease subtype to the disease-free individuals \citep{wang2016statistical}, then $\exp({\beta}_k)$ is the OR analogue of \eqref{Eq:NaiveCond}, and approximates the RR as CRC is a rare disease. However, \eqref{Eq:NaiveCond} does not correspond to a causal effect,   because of selection bias due to  common causes of the different subtypes \citep{nevo2021reflection}. These common causes cannot be classified as classical confounders, as they can be independent of the exposure. If these common causes are also mediators of an exposure-subtype mechanisms that are shared between more than one subtype, even had they been observed, adjusting for them would have eliminated part of the causal effect of the exposure on the subtype  \citep{nevo2021reflection}. Importantly, because MSI-high and non-MSI-high are subtypes of  the same disease, CRC, they are likely to have  unknown and/or unmeasured common causes.

\cite{young2020causal} considered possible causal estimands in the presence of competing events. The \textit{total effect}, 
 $TE^{(1)} = \bbE[Y^{1}(1) - Y^{1}(0)], $
contrasts the population-level subtype $1$ risk had the entire population been exposed, with the population-level subtype $1$ risk had the entire population were unexposed.  In the presence of measured confounders, this effect is identified from the observed data under SUTVA, weak ignorability and positivity by
\begin{equation}
\label{Eq:TEidentStand}
TE^{(1)} = \bbE_{\bX}[\bbE(Y^{(1)} | A=1, \bX)] - \bbE_{\bX}[\bbE(Y^{(1)} | A = 0,\bX)].
\end{equation}
 Unlike \eqref{Eq:NaiveCond}, the definition of $TE^{(1)}$  does not involve conditioning on the post-exposure variable of not having the other subtypes.  Note also that if model \eqref{Eq:SunModel} is correctly specified, $\exp(\widetilde{\beta}_k)$ is the total effect for subtype $k$ on the OR/RR scales.
 
 However, despite $TE^{(1)}$ being a valid causal effect, it does not suffice to describe exposure effects in the presence of etiologic disease heterogeneity. The total effect of $A$ on $Y^{(1)}$, for example, includes the effect caused by changes in $Y^{(k)}$ , $k>1$, or, in other words, $TE^{(1)}$ is sensitive to the phenomenon of exposure-induced subtype switching. For example, assume some individuals would have been diagnosed with subtype 1 under $A=0$ and would have been diagnosed with subtype 2 under $A=1$, while the rest of the population will be free of subtype 1 for all $A$ values. Then, the total effect on subtype 1 will be negative,  $TE^{(1)}<0$, and it would appear as if the exposure protects from subtype 1, even though the individuals ``protected'' by $A$ would still have the disease, just of another subtype.  

The \textit{direct effect} \citep{young2020causal} is the causal effect comparing subtype 1 rates under two hypothetical joint interventions. The first sets the exposure to $A=1$ and eliminates the possibility of disease subtypes other than subtype 1 disease, and the second sets the exposure to $A=0$ and also eliminates risks of other disease subtypes. This approach is, however, irrelevant for our  study, because an intervention eliminating the risk of a particular CRC subtype is not currently available and is not expected to be in the near future.

Another alternative considers the \textit{separable effects} \citep{stensrud2020separable}, which rely on a conceptual separation of the exposure into the components affecting each disease subtype. Such separation requires profound understanding of the mechanisms leading to each CRC subtype.   Furthermore, the conditions  needed for identifiability of these effects are violated if there are unobserved common causes of both disease subtypes \citep{stensrud2020separable}. Such common causes are expected to exist in our study and in many disease subtype studies \citep{nevo2021reflection}.


\section{The Subtype-Free Average Causal Effect}	\label{Sec:SFACE}	
	
Our focus in this paper is on the common scenario of two disease subtypes ($K=2$), as in our study. We propose a new estimand, inspired by the $SACE$ \citep{rubin2006causal}. We define the $\sface$ on the difference scale by
\begin{equation}
\label{Eq:SFACEdef}
\sface^{(1)}_{D} = \mathbb{E}[Y^{(1)}(1)- Y^{(1)}(0)\ |\ Y^{(2)}(0) = Y^{(2)}(1) = 0],
\end{equation}
and analogously define
\begin{equation*}
\sface^{(1)}_{RR} 
= \frac {\bbE[Y^{(1)}(1)| Y^{(2)}(0) = Y^{(2)}(1) = 0]} {\bbE[Y^{(1)}(0) | Y^{(2)}(0) = Y^{(2)}(1) = 0]}
\end{equation*}
to be the $\sface$ on the RR scale. 
The $\sface^{(1)}$ is the average causal effect among the individuals that would have been free of disease subtype 2 under either exposure level (i.e., regardless of smoking status). 
Similarly to $\sface^{(1)}$, let
$$ 
\sface^{(2)}_D = \mathbb{E}[Y^{(2)}(1)- Y^{(2)}(0) \ | \  Y^{(1)}(0) = Y^{(1)}(1) = 0],
$$ 
be the average causal effect on the difference scale of $A$ on $Y^{(2)}$ among the individuals that would have been free of disease subtype 1 under either exposure level. Let also $\sface^{(2)}_{RR}$ be the analogous causal effect on the RR scale. Disease heterogeneity studies often focus on whether the exposure has the same effect on both disease subtypes. For investigating the heterogeneity of the causal effects, one can use $\theta_D = \sface^{(1)}_D - \sface^{(2)}_D$ and test the null hypothesis of $H_0: \theta_D = 0$ versus the alternative of $H_1: \theta_D \ne 0$. 

\subsection{Interpretation of the $\sface$}
\label{SubSec:Interpret}

As explained in Section \ref{SubSec:ReviewCausal}, competing events present a challenge in defining causal estimands that shed light on the scientific question of interest. The $\sface$ presents another source of information. A positive $\sface^{(1)}_D$ value means that in the sub-population who would have been free of MSI-high CRC regardless of their smoking status, we will expect more non-MSI-high CRC cases had the entire sub-population smoked compared to the scenario the entire sub-population did not smoke before age 60. 

Therefore, the $\sface$ offers a way to circumvent non-zero causal effects due to exposure-induced subtype switching. Nevertheless, the above appealing advantage does not come without a price. One disadvantage is that the $\sface$ is defined in a latent subset of the population. This is a general problem with principal stratification approaches. A second, related, issue is that when contrasting  $\sface^{(1)}$ and $\sface^{(2)}$, we contrast effects defined in two overlapping, but not identical, sub-populations. The $\sface^{(1)}$ is defined within the sub-population $\{Y^{(2)}(0) = Y^{(2)}(1) = 0\}$, and the $\sface^{(2)}$ is defined within the sub-population $\{Y^{(1)}(0) = Y^{(1)}(1) = 0\}$. 

Thus, a non-zero $\theta_D$ captures changes that might be due to the different population, and as a result, the exact source of heterogeneity in the causal effect is not revealed. Instead, a non-zero $\theta_D$ presents evidence of heterogeneity and calls for further investigation. Lastly but importantly, in rare disease scenarios, such as CRC, this problem is less concerning because the two sub-populations $\{Y^{(2)}(0) = Y^{(2)}(1) = 0\}$ and $\{Y^{(1)}(0) = Y^{(1)}(1) = 0\}$ have almost perfect overlap, as those who would not have CRC of any subtype under any exposure value belong to both groups and comprise the vast majority of the population.

\subsection{Identifiability}	
\label{SubSec:Identifiability}

Similar to the $SACE$ in truncation-by-death scenarios, the $\sface$ is not identifiable from the data under the standard assumptions of SUTVA, weak ignorability and positivity. One commonly-made assumption when targeting the SACE is Monotonicity \citep{zhang2003estimation}, stating that treatment cannot hurt survival.  In our study, the definition and plausibility of Monotonicity-like assumptions are more nuanced, and, as it turns out, resulting in new insights regarding principal causal effects
and their identification in a competing risks setup.  We start with the definition of \textit{Subtype Monotonicity} (\SmonoNO).
\begin{assumption} 
Subtype Monotonicity (\SmonoNO): \label{Ass:Smono}
$Y_i^{(k)}(0) \le Y_i^{(k)}(1)$. 
\end{assumption}
\Smono states that an individual that would have been diagnosed with disease subtype $k$ under no exposure (never smoking), would have also been diagnosed with disease subtype $k$ under exposure (ever smoking). Therefore, an individual who would have been free of disease subtype $k$ under exposure, would have also been free of disease subtype $k$ under no exposure. In other words, \Smono asserts that the exposure cannot prevent the subtype $k$-specific disease. \Smono trivially holds if the exposure does not affect the outcome $Y$ at all, and can be easily modified to address scenarios in which the exposure is suspected to have a protective effect.

Importantly, \Smono is a subtype-specific assumption. It can hold for both subtypes, for one subtype, or for none of them. Interestingly, we show in Section A of the Appendix that under \Smono for both subtypes, the $\sface^{(k)}_{RR}$ equals to the total effect $TE^{(k)}_{RR}$. When the effects are defined on the difference scale, this equality does not hold.  

It is clear from the above that the $\sface_{RR}$ is identifiable from the data under SUTVA, weak ignorability, positivity, and \Smono for both subtypes.  The following proposition establishes identification of $\sface_D$, and presents identification formulas under either standardization or IPTW \citep{hernan2020causal}.
\begin{proposition}
\label{prop:SmonoIdent}
Under SUTVA, weak ignorability, positivity, and \Smono for subtypes $k=1,2$, the $\sface^{(1)}_D$ is identifiable from the observed data by
\begin{align}
\label{Eq:sfaceStand}
\sface^{(1)}_{D} &=  \frac{\bbE_{\bX}\left[\pi_1(1,\bX)\right] - \bbE_{\bX}\left[\pi_1(0,\bX)\right]}{ 1-\bbE_{\bX}\left[\pi_2(1,\bX)\right]},\\ \nonumber
& \hspace{-6cm} \text{using standardization, or via IPTW by}\\
\label{Eq:sfaceIPTW}
\sface^{(1)}_{D} &= \frac{\bbE\left[ \frac{\bbmI \{A=1\} Y^{(1)}}{e(\bX)}\right] - \bbE\left[ \frac{ \mathbbm{1} \{A=0\} Y^{(1)}}{1-e(\bX)}\right]}{1 - \bbE\left[ \frac{ \mathbbm{1} \{A=1\} Y^{(2)}}{e(\bX)}\right]}.
\end{align}
\end{proposition}
The proofs and analogous expressions for $\sface^{(2)}$,  $\theta_{D}$, and for causal contrasts on the RR scale are all given in Section A of the Appendix.  Estimation based on these identification formulas is discussed in Section \ref{SubSec:EstInfer}.

\subsection{Relaxing \SmonoNO}
\label{SubSec:RelaxingMono}

Assuming \Smono for both subtypes might be too restrictive, even if we believe the exposure cannot protect from the outcome, as \Smono disallows subtype switching for any individual in the population. For example, consider the scenario the exposure is believed to increase non-MSI-high CRC risk but not MSI-high CRC risk. If individual $i$ would have been diagnosed with MSI-high CRC when unexposed, then under \Smono for MSI-high CRC, it will be impossible for individual $i$ to be diagnosed with non-MSI-high CRC when exposed. This is even though the exposure is a risk factor only for non-MSI-high CRC. 

Therefore, we consider a weaker assumption, termed \textit{Disease Monotonicity} (\DmonoNO), that states that an individual who would have been diagnosed with disease subtype $k$ under no exposure, would have also been diagnosed with the disease under exposure, but not necessarily with the same subtype. 
\begin{assumption} Disease Monotonicity (\DmonoNO): $Y_i^{(k)}(0) \le \max\{Y_i^{(1)}(1),Y_i^{(2)}(1)\}$.
\label{Ass:PartSmono}
\vspace{-1cm}
\end{assumption}
\Dmono  keeps the premise that the exposure cannot protect from the disease while being less restrictive than S-Monotonicity.  As with \SmonoNO, \Dmono can be modified for scenarios where the exposure is suspected to have a protective effect.
	
For each disease subtype, we may assume \SmonoNO, \DmonoNO, or neither. With two disease subtypes, we have eight different combinations of possible assumptions. Under each combination, certain potential outcome profiles of $\{Y_i^{(1)}(0),Y_i^{(1)}(1),Y_i^{(2)}(0),Y_i^{(2)}(1)\}$ are assumed to be absent from the population. For example, under \Smono for subtype 1, there is no one in the population with $\{Y_i^{(1)}(0),Y_i^{(1)}(1),Y_i^{(2)}(0),Y_i^{(2)}(1)\}= \{1,0,0,1\}$. Note also that because the subtypes are mutually exclusive, it is impossible to have $Y_i^{(k)}(0)=Y_i^{(k)}(1)=1$ regardless of the taken monotonicity assumptions.

Table \ref{Tab:aspo} presents the nine possible potential outcome profiles, and specifies whether each profile exists under four different combinations of assumptions. This table demonstrates that the exact implications of the different monotonicity assumptions on the potential outcomes in the population are quite explicit. The observed data, however, does not  hold all the information required for classifying each individual into their potential outcome profile. Table A.1 in Section A of the Appendix specifies the possible profiles for each participant according to their observed $(A, Y^{(1)}, Y^{(2)})$  under each of the four combinations of assumptions. 

\begin{table}[ht]
\captionsetup{font=footnotesize}
\centering
\footnotesize{
\resizebox{\textwidth}{!}{
\begin{tabular}{ P{1cm}P{5cm}P{2.5cm}P{2.5cm}P{2.5cm}P{2.5cm}}
\hline
\multicolumn{2}{c}{Potential outcomes profiles} & \multicolumn{4}{c}{Monotonicity assumptions} \\
\hline
Profile & $\{Y^{(1)}(0), Y^{(1)}(1), Y^{(2)}(0),Y^{(2)}(1)\}$ & 
\thead{$Y^{(1)}$ S-Mono\\$Y^{(2)}$ S-Mono} & 
\thead{$Y^{(1)}$ D-Mono\\$Y^{(2)}$ D-Mono} & 
\thead{$Y^{(1)}$ S-Mono\\$Y^{(2)}$ D-Mono} & 
\thead{$Y^{(1)}$ D-Mono\\$Y^{(2)}$ S-Mono} \\
\hline
0 & $\{0, 0, 0, 0\}$  & \cmark & \cmark & \cmark & \cmark  \\ 
1 & $\{0, 0, 0, 1\}$   & \cmark & \cmark & \cmark & \cmark  \\
2 & $\{0, 0, 1, 0\}$ & \xmark & \xmark & \xmark & \xmark  \\
3 & $\{0, 1, 0, 0\}$   & \cmark & \cmark & \cmark & \cmark   \\
4 & $\{1, 0, 0, 0\}$   & \xmark & \xmark & \xmark & \xmark  \\
5 & $\{1, 1, 0, 0\}$ & \cmark & \cmark & \cmark & \cmark   \\
6 & $\{0, 0, 1, 1\}$  & \cmark & \cmark & \cmark & \cmark \\			
7 & $\{1, 0, 0, 1\}$ & \xmark & \cmark & \xmark & \cmark  \\
8 & 
$\{0, 1, 1, 0\}$ & \xmark & \cmark & \cmark & \xmark  \\
\hline
\end{tabular}}}
\caption{Nine possible potential outcome profiles and their (im)possibility under different assumption combinations. Each row represents one profile, with \xmark $ $ indicating assumption combinations under which no individual in the population have this profile, and \cmark $ $ indicating this profile may exists in the population.}
\label{Tab:aspo}
\end{table}

\subsection{A sensitivity analysis approach}
\label{SubSec:Sens}
By Proposition \ref{prop:SmonoIdent}, under \Smono for both disease subtypes, the $\sface^{(1)}$, $\sface^{(2)}$ and $\theta$ are identifiable from the observed data distribution. If we replace \Smono with \Dmono  or with no monotonicity assumption for at least one subtype, then the causal effects are no longer  identifiable, and  $\sface^{(k)}_{RR}$ does not equal to the $TE^{(k)}_{RR}$.  

Therefore, we take a sensitivity analysis approach, as is common when studying causal effects under untestable assumptions. Let $\lambda_1 = \Pr[Y^{(2)}(1) = 1|Y^{(1)}(0) = 1]$ be the proportion of individuals who would have been diagnosed with disease subtype 2 when exposed, out of the individuals who would have been diagnosed with disease subtype 1 when unexposed. That is, $\lambda_1$ is the probability of exposure-induced subtype switching from subtype 1 to subtype 2. Note that $\lambda_1=0$ under \Smono for subtype 1 and   $\lambda_1\ge 0$ under \Dmono  for subtype 1.   Let also $\lambda_2 = \Pr[Y^{(1)}(1) = 1|Y^{(2)}(0) = 1]$ be the subtype-switching probability from subtype 2 to subtype 1. While $\lambda_1$ and $\lambda_2$ are not identifiable from the observed data, they can be bounded. Specifically, because $\lambda_1 \le \min \left\{ 1, \frac{\Pr(Y^{(1)}(1) = 1)}{\Pr(Y^{(2)}(0) = 1)} \right\}$, a data-driven   upper bound for $\lambda_1$ 
is $\min \left\{1, \frac{\bbE_{\bX}[\pi_1(1,\bX)]}{\bbE_{\bX}[\pi_2(0,\bX)]}\right\}$. Similarly,  
$\lambda_2 \le \min \left\{ 1, \frac{\bbE_{\bX}[\pi_2(1,\bX)]}{\bbE_{\bX}[\pi_1(0,\bX)]} \right\}$.


The following two propositions demonstrate that the subtype-switching probabilities play a key role in the identification of the $\sface$ whenever the \Smono assumptions are not met. Proposition \ref{prop:DmonoIdentDiff} shows that (a) under \Dmono  for subtype 1 and \Smono for subtype 2, both the $\sface^{(1)}_D$ and $\sface^{(2)}_D$ are identifiable as functions of $\lambda_1$, (b) under \Dmono  for subtype 2 and \Smono for subtype 1, the $\sface^{(1)}_D$ and $\sface^{(2)}_D$ are identifiable as functions of $\lambda_2$, and (c) under \Dmono for both subtypes, the $\sface$s are identifiable  as functions of $\lambda_1$ and $\lambda_2$.

\begin{proposition}
\label{prop:DmonoIdentDiff}
Under SUTVA, weak ignorability, positivity and \Dmono  for both subtypes, the $\sface^{(1)}_D$ is identifiable from the data by 
\begin{align}
\label{Eq:sface1StandLambdas}
\sface^{(1)}_D &= \frac{\bbE_{\bX}[\pi_1(1,\bX)] +(\lambda_1 - 1) \bbE_{\bX}[\pi_1(0,\bX)] - \lambda_2  \bbE_{\bX}[\pi_2(0,\bX)]} { 1-\bbE_{\bX}[\pi_2(1,\bX)] - \lambda_2  \bbE_{\bX}[\pi_2(0,\bX)]} ,\\[1em] \nonumber
& \hspace{-3.7cm} \text{using standardization, or via IPTW by}\\
\label{Eq:sface1IPTWLambdas}
\sface^{(1)}_{D} &= \frac{\bbE\left[ \frac{\bbmI \{A=1\} Y^{(1)}}{e(\bX)}\right] + (\lambda_1-1) \bbE\left[ \frac{ \bbmI \{A=0\} Y^{(1)}}{1-e(\bX)}\right]-\lambda_2\bbE\left[\frac{\mathbbm{1}\{A=0\}Y^{(2)}}{1-e(\bX)}\right]}{1 - \bbE\left[\frac{\bbmI \{A=1\} Y^{(2)}}{e(\bX)}\right]-\lambda_2\bbE\left[\frac{\bbmI\{A=0\}Y^{(2)}}{1-e(\bX)}\right]}.
\end{align}
The identification of $\sface^{(2)}_D$ is analogous. 
\end{proposition}
The proof is given in Section A of the Appendix. Of note is that if one further assumes \Smono for one of the subtypes (say subtype 2),  then an  identification formula of $\sface^{(1)}_D$ is obtained by setting the appropriate $\lambda_k$ (say $\lambda_2$) to zero in \eqref{Eq:sface1StandLambdas} and \eqref{Eq:sface1IPTWLambdas}. A second remark is that, technically, results \eqref{Eq:sface1StandLambdas} and \eqref{Eq:sface1IPTWLambdas} require \Dmono  for subtype 2 only, and similarly, the  identification for subtype 2 requires \Dmono  for subtype 1 only. In Section A of the Appendix, we present further identification results for $\sface_D$ under \Dmono for one subtype only or without any monotonicity assumptions, and show that in these cases $\lambda_1$ and $\lambda_2$ do not suffice for identification of $\sface_D$. 

Turning to the RR scale, Proposition \ref{prop:DmonoIdentRR} asserts that  $\sface^{(k)}_{RR}$ is identifiable as a function of $\lambda_1$ and $\lambda_2$ and the observed data without imposing any monotonicity assumptions. 
\begin{proposition}
\label{prop:DmonoIdentRR}
Under SUTVA, weak ignorability and positivity, the $\sface^{(1)}_{RR}$ is identifiable via standardization by
\begin{equation*}
    \sface^{(1)}_{RR}=   \frac{\bbE_{\bX}[\pi_1(1,\bX)] - \lambda_2  \bbE_{\bX}[\pi_2(0,\bX)]} {(1-\lambda_1) \bbE_{\bX}[\pi_1(0,\bX)] },
\end{equation*}
or using IPTW, 
\begin{equation*}
\sface^{(1)}_{RR}
 =\frac{\bbE\left[ \frac{\bbmI \{A=1\} Y^{(1)}}{e(\bX)}\right]-\lambda_2\bbE\left[\frac{\mathbbm{1}\{A=0\}Y^{(2)}}{1-e(\bX)}\right]}{(1-\lambda_1) \bbE\left[ \frac{ \bbmI \{A=0\} Y^{(1)}}{1-e(\bX)}\right]}, 
\end{equation*}
The identification of $\sface^{(2)}_{RR}$ is analogous. 
\end{proposition}
The proof is given in Section A of the Appendix. 

Propositions \ref{prop:DmonoIdentDiff} and \ref{prop:DmonoIdentRR} serve as  basis for our proposed sensitivity analyses. Researchers first decide what is the effect scale of interest, and then for which if any of the two subtypes the \Smono or \Dmono assumptions are plausible. Then, depending on the taken assumptions, they may set $\lambda_1$ or $\lambda_2$ to zero, and estimate the $\sface^{(k)}$ and $\theta$ as a function of $\lambda_1$ and/or $\lambda_2$. Such an analysis can reveal how sensitive the results are to different levels of exposure-induced subtype switching. We implement these analyses in our study of smoking effect on CRC MSI subtypes in Section \ref{Sec:DataAnalysis}.

\subsection{Estimation}
\label{SubSec:EstInfer}

Recall that for each individual $i$, the observed data are ($A_i, \bX_i, Y_i$) or equivalently  ($A_i, \bX_i, Y^{(1)}_i, Y^{(2)}_i$). Estimation of $\sface^{(k)}$ and $\theta$ from the data alongside the sensitivity analyses for the \Smono assumptions require estimation of $\pi_k(a,\bx)$, $k=1,2,$ $a=0,1$ for the standardization-based formulas, or $e(\bx)$ for the IPTW-based identification formulas. Expectations over $\bX$ or $(\bX,A,Y^{(k)})$ are estimated by taking the sample analogues.

Non-parametric estimation of $\pi_k(a,\bx)$ or $e(\bx)$ when the vector of measured confounders $\bX$ is of high dimension or includes many continuous variables could be a difficult task, that will yield an estimator with a too large variance to be able to derive conclusions from the data. Therefore, parametric estimation might be a better option, and also aligns with the models used in practice and that were presented in Section \ref{SubSec:ReviewStat}. We first discuss parametric estimation, before implementing a DR estimator that will be consistent even if not all models are correctly-specified.

Starting with standardization, let $\bpi(a,\bx;\beeta_1)=(\pi_0(a,\bx;\beeta_1),\pi_1(a,\bx;\beeta_1),\pi_2(a,\bx;\beeta_1))$ be the parametric model for $\bpi(a,\bX)$ with parameter vector $\beeta_1$. This could be,  for example, the regression models \eqref{Eq:MultinoModel} or   \eqref{Eq:SunModel}. Upon estimating $\hat{\beeta}_1$,  the estimated probabilities $\bpi(a,\bX_i;\hat{\beeta}_1)$ are  calculated in the entire sample, for $a=0,1$, before plugged in the relevant identification formulas. For example, the proposed $\sface^{(1)}_D$ estimator  is
\begin{equation}
\label{Eq:standest}
\widehat{\sface^{(1)}}_{D,stand} =  \frac{ \sum_{i=1}^n \pi_1(1,\bX_i;\hat{\beeta}_1)  -  \sum_{i=1}^n \pi_1(0,\bX_i;\hat{\beeta}_1)}{ 1-\sum_{i=1}^n \pi_2(1,\bX_i;\hat{\beeta}_1)}.
\end{equation}
If there is an unmeasured covariate $U$ affecting both disease subtypes $(Y^{(1)}, Y^{(2)})$, model \eqref{Eq:MultinoModel} might be misspecified, if one believes the model that is also conditioned on $U$ is a multinomial regression model. However, this is not a main concern in studies involving rare  disease, as in our study \citep{nevo2021reflection}. 

Turning to IPTW, let  $e(\bx;\beeta_2)$ be a parametric model for $e(\bx)$, with parameter vector $\beeta_2$, for example a logistic regression. Then, after estimating $\hat{\beeta}_2$, the $\sface^{(1)}_D$ estimator is obtained by 
\begin{equation}
\widehat{\sface^{(1)}}_{D, IPTW} = \frac{ \mathlarger{\sum}_{i=1}^n {\frac{A_i Y^{(1)}_i }{e(\bX_i,\hat{\beeta}_2)}} - \mathlarger{\sum}_{i=1}^n \frac{(1-A_i) Y^{(1)}_i }{e(\bX_i,\hat{\beeta}_2)}}{ 1-\mathlarger{\sum}_{i=1}^n \frac{A_i Y^{(2)}_i }{e(\bX_i,\hat{\beeta}_2)}},
\label{eqn:iptwest}
\end{equation}
and similar estimators for $\sface^{(2)}_D$ and $\theta_D$, or for effects on other scales, can be obtained. 

When using the maximum likelihood approach to fit models $\bpi(a,\bx;\beeta_1)$ or $e(\bx,\beeta_2)$, the obtained estimators $\widehat{\beeta}_1$ or $\widehat{\beeta}_2$ are consistent and asymptotically normal. Thus, the continuous mapping theorem and the delta method  can be used to show that the $\sface$ and $\theta$ estimators are also consistent and asymptotically normal. In practice, the bootstrap can be used to estimate the standard errors (SEs) and construct confidence intervals (CIs).

The standardization estimator or the IPTW estimator will be consistent only if their respective models are correctly specified. Therefore, we extend our estimation approach to DR estimation \citep{robins1994estimation,lunceford2004stratification}. That is, we construct an estimator that will be consistent whenever at least one of the two models, $\bpi(a,\bx;\beeta_1)$ or $e(\bx;\beeta_2)$, but not necessarily both, is correctly specified. DR estimators are frequently used in causal inference to reduce dependency on the chosen model. Furthermore, a DR estimator constructed as an augmented-IPTW estimator is also more efficient than IPTW estimators and have desirable theoretical properties \citep{robins1994estimation}.

Let $\hat{\beeta}_1$ and $\hat{\beeta}_2$ be the estimators obtained for $\beeta_1$ and $\beeta_2$. As can be seen from the proof of Proposition \ref{prop:SmonoIdent} (Section A of the Appendix), under \SmonoNO, the $\sface$s and $\theta$ are functions of $\Pr[Y^{(k)}(a)=1], a=0,1,$ $k=1,2$. For example, $\sface^{(1)}$ is equal to $\frac{\Pr[Y^{(1)}(1) = 1] - \Pr[Y^{(1)}(0) = 1]}{\Pr[Y^{(2)}(1) = 0]}$.
Under weak ignorability and positivity, a DR augmented-IPTW estimator for $\Pr[Y^{(k)}(1)=1]$ is 
\begin{equation}
\label{Eq:Yk1DR}
\frac{1}{n}\sum_{i=1}^n\left[\frac{A_iY_i^{(k)}}{e(\bX_i;\hat{\beeta}_2)} - \frac{[A_i-e(\bX_i;\hat{\beeta}_2)]\frac{1}{n}\sum_{j=1}^n\pi_k(1,\bX_j;\hat{\beeta}_1)}{e(\bX_i;\hat{\beeta}_2)} \right]
\end{equation}
\citep{bang2005doubly,lunceford2004stratification}. The estimator \eqref{Eq:Yk1DR} and analogous expressions for $\Pr[Y^{(k)}(0)=1]$ can be then plugged in $\frac{\Pr[Y^{(1)}(1) = 1] - \Pr[Y^{(1)}(0) = 1]}{\Pr[Y^{(2)}(1) = 0]}$ 
and similar equations to obtain estimators for the causal effects that are consistent by the continuous mapping theorem, if at least one of the models, $\pi(a,\bx;\beeta_1)$ or $e(\bx;\beeta_2)$, is correctly specified. Similar arguments can be used to construct DR estimators when conducting sensitivity analyses based on Proposition \ref{prop:DmonoIdentDiff} and \ref{prop:DmonoIdentRR}. 
 

\section{Simulations}
\label{Sec:Sims}

We conducted simulation studies to evaluate the finite-sample performance of our proposed estimators and to compare the $\sface$ to  existing estimands. 
Detailed information about the data generating mechanism and parameters used, as well as more detailed results are given in Section B of the Appendix. For each $i=1,...n$, potential outcomes $Y_i(a)$ were simulated using a multinomial regression model with two confounders, $X_{1i} \sim Ber(0.5)$, $X_{2i}\sim N(0,1)$, and one unobserved covariate $U_i\sim N(0,1)$ affecting both subtypes. \Smono was imposed for both subtypes (without violating the multinomial model assumption, see Section B). The exposure $A_i$ was simulated using a logistic regression with $X_{1i}$ and $X_{2i}$. The observed outcomes $Y_i$ were obtained by $Y_i = A_i Y_i(1) + (1-A_i) Y_i(0)$. 

To study different aspects of the proposed approach, we conducted three studies. In the first, we fixed the model parameters and varied the sample size between 5,000 and 50,000, to study the methods' performance in different sample sizes. In the second study, we fixed the sample size and varied the effect the unmeasured covariate $U_i$ had on $Y_i^{(2)}$ to compare between the different causal effects. In the third study, we investigated the methods' performance when the fitted models were misspecified. 

In each simulated dataset, we first fitted the multinomial regression \eqref{Eq:MultinoModel}, and a logistic regression for $e(\bx)$. Then, we calculated the conditional estimators \eqref{Eq:NaiveCond}, the TE estimators \eqref{Eq:TEidentStand}, and the three $\sface$ estimators described in Section \ref{SubSec:EstInfer}. We used the bootstrap with 200 repetitions to estimate the SEs and to construct Wald-type 95\% CIs.

In the first study (Tables B.3 and B.4),  the $\sface$ estimators had a small relative bias. The SE of all estimators were similar, and were generally well-estimated. Empirical coverage rates of the CIs were satisfactory. These results were also apparent for the second study (Figures B.1 and B.2).  In the second study, as the effect of $U$ on $Y^{(2)}$ increased, absolute bias of the conditional estimand and the TE on the difference scale, compared to the true $\sface$s,   increased, while the $\sface$ estimators remained unbiased. As expected, because data were simulated under \Smono for both subtypes, the $TE^{(k)}_{RR}$ and $\sface^{(k)}_D$ coincided. Finally,  the last study results demonstrated that when at least one of the models, $\bpi(a,\bx;\beeta_1)$ or $e(\bx)$ is correct, the DR estimator was the only estimator with negligible bias (Tables B.5 and B.6).  

\section{Data analysis}
\label{Sec:DataAnalysis}

We conducted several analyses to study the causal effects of ever smoking by age 60 on having CRC MSI subtypes by age 70, using the NHS and HPFS datasets described in Section \ref{Sec:DataDesc}. The main results are presented here, and additional results are given in Section C of the Appendix. In all of our analyses, we used the confounders detailed in Section \ref{Sec:DataDesc} and Table C.7.  Wald-type 95\% CIs and $p$-values were calculated using the bootstrap with 200 repetitions. 

As indicated in Section \ref{Sec:DataDesc}, for more than 50\% of CRC cases the subtype status was unknown. To minimize the impact of potential selection bias, we used inverse probability weighting for missing subtype \citep{liu2018utility}. We first fitted a model for the probability of observing the subtypes (Table C.8). The model included additional variables available for CRC cases (e.g., tumor stage and size). Then, for each CRC case with known subtype, we calculated the weight as the reciprocal of the probability of observing the subtype. The weights were truncated at the 99\% percentile and then incorporated into our estimators.

We estimated the TE of smoking on each disease subtype by standardization \eqref{Eq:TEidentStand}, using a multinomial regression model \eqref{Eq:MultinoModel}  (Table C.9). The point estimates indicated smoking increases the  risk of both CRC MSI subtypes  (Tables \ref{Tab:DataRes} and C.10), although the results were not significant at the 5\% level. The point estimates indicated that the effect on non-MSI-high subtype was larger than the effect on the MSI-high subtype  on the difference scale and vice versa on the RR scale. These results are in agreement with the existing literature, because MSI-high is more rare, so while the RR is larger, the number of excess cases due to smoking can be smaller. 

\begin{table}[!htbp]
\centering
\small
\begin{tabular}{llllll}
Subtype & Effect & Method & Estimate & $\widehat{SE}$ & 95\%CI \\ 
  \hline
non-MSI-high  & $\sface_D$ & Stand & 129.9 & 87.6  & [-41.8, 301.6] \\  & &IPTW &  141.8 & 89.7 & [-34.0, 317.6] \\
 & &DR & 161.3 & 77.9 & [ 8.6, 314.0] \\ 
 & $TE_D$ & Stand & 129.8 & 88.6 & [-43.9, 303.5] \\ 
 MSI-high & $\sface_D$ & Stand & 32.7 & 29.8 & [-25.7, 91.1] \\ 
 & & IPTW & 34.8 & 26.3 &  [-16.7, 86.3] \\ 
  & & DR & 37.2 & 23.9 & [-9.6, 84.0] \\ 
& $TE_D$ & Stand & 32.4 & 28.5 & [-23.5, 88.3] \\ 
\end{tabular}
\captionsetup{font=footnotesize}
\caption{Estimated $\sface_D$  and $TE_D$ of smoking  on the two CRC subtypes under \Smono for both subtypes. Effects presented per 100,000 people. Stand: standardization; $\widehat{SE}$: estimated standard error; 95\%CI: 95\% confidence interval.}
\label{Tab:DataRes}
\end{table}

Turning to the $\sface s$, we first considered an analysis under \Smono for both disease subtypes. For the standardization-based estimator we used the multinomial regression model \eqref{Eq:MultinoModel}. For the IPTW-based estimator  we used a logistic regression for the exposure (Table C.10). We used both models to calculate the DR estimator. As expected, for causal effects defined on the RR scale, the standardization-based $\sface$ and TE estimators were the same (Table C.11). Compared with the TE, the $\sface s$ on the difference scale (Table \ref{Tab:DataRes})  indicated larger causal effects for both subtypes within
the appropriate principal strata. The estimated effects were the largest using the DR estimators, which also had the lowest estimated SE. The only significant smoking $\sface$ effect was obtained by the DR estimator for the  non-MSI-high subtype, who estimated additional 161.3 cases per 100,000 people (95\%CI: 8.6, 314.0), had the entire principal stratum free of MSI-high-subtype been smokers (compared with had they were all non-smokers). The RR estimate was 1.26 (95\%CI: 1.03, 1.58). The estimated effect on MSI-high subtype was smaller on the difference scale (37.2 additional MSI-high cases per 100,000 people, 95\%CI: -9.6, 84.0) and larger on the RR scale (1.48, 95CI\%: 0.64, 2.32) than the corresponding effects on the non-MSI-high subtype. Nevertheless, the null hypothesis of $H_0: \theta_D = 0$ was not rejected ($p=0.148$).

\begin{figure}
\centering
\includegraphics[scale=0.3]{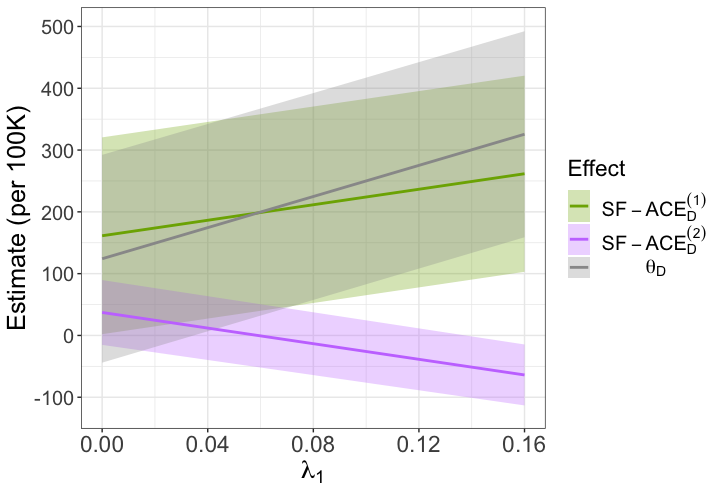}
\captionsetup{font=footnotesize}
\caption{Sensitivity analysis for causal effects under \Smono for MSI-high (subtype 2) and \Dmono  for non-MSI-high  (subtype 1) CRC subtypes. The figure presents DR estimates as a function of $\lambda_1$. The shadows are 95\% Wald-type CIs connected continuously for clarity of presentation.}
\label{Fig:SensDmono1}
\end{figure}

The \Smono assumption might be too restrictive, especially for the non-MSI-high subtype, for which there is limited information about the relative effect strength \citep{Amitay2020smoking}. From a subject-manner perspective, we could not overlook the possibility of exposure-induced subtype switching for non-MSI-high subtype, as the relative effect of smoking on MSI-high CRC is known to be larger. Therefore, we replaced the \Smono assumption for non-MSI-high (subtype 1) with \Dmono  by fixing $\lambda_2=0$  and varying the values of $\lambda_1$ between 0 and $ \frac{n^{-1}\sum_{i=1}^{n}[\widehat{\pi}_1(1,\bX_i)]}{n^{-1}\sum_{i=1}^{n}[\widehat{\pi}_2(0,\bX_i)]}=0.17$ (Figure \ref{Fig:SensDmono1}). As $\lambda_1$ increased, the estimated $\sface$ for non-MSI-high CRC increased, the estimated $\sface$  for MSI-high CRC decreased. For $\lambda_1\ge 0.04$,  $H_0: \theta_D = 0$ was rejected. This result serves as evidence for heterogeneity in the $\sface$  as long as we believe subtype switching is possible and not extremely rare.

\begin{figure}
\centering
\includegraphics[scale=0.45]{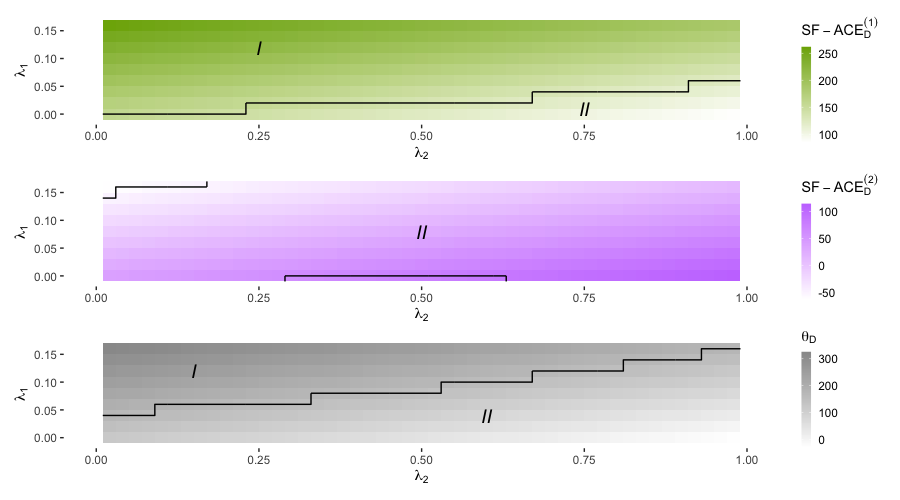}
\captionsetup{font=footnotesize}
\caption{Sensitivity analysis for estimating $\sface_D^{(1)}$, $\sface_D^{(2)}$ and $\theta_{D}$ under \Dmono  for both  CRC MSI subtypes. The figure presents DR estimates as a function of $\lambda_1= \Pr[Y^{(2)}(1) = 1|Y^{(1)}(0) = 1]$ and $\lambda_2 = \Pr[Y^{(1)}(1) = 1|Y^{(2)}(0) = 1]$. The black lines divide the grid such that in sections marked with $I$ the effect is significant in 5\% level while in sections marked with $II$ it is not.}
\label{Fig:SensDmonoBoth}
\end{figure}

Given some a-priori evidence for smoking effects on both subtypes, we also considered an analysis under \Dmono  for both subtypes. Figure \ref{Fig:SensDmonoBoth} presents the estimated effects on the difference scale as a function of $\lambda_1$ and $\lambda_2$ (for which the data did not restrict the range), along with 95\% CIs. For nearly the entire range, the effect on non-MSI-high CRC was significant and the effect on MSI-high CRC was not. Sensitivity analysis results were similar for all three estimators (Figures \ref{Fig:SensDmonoBoth}, C.4 and C.5). As $\lambda_2$ increased, the larger was the minimal value of $\lambda_1$ for which  heterogeneity can be concluded. For the RR (Figures C.6--C.8),  under the combination of  low $\lambda_2$ and large $\lambda_1$ the heterogeneity assumption was rejected.   
	
As described in Section \ref{Sec:DataDesc}, there were 17,854 study participants for which the last available data were before age 70 (and they were not diagnosed with CRC and did not die). We treated these people as CRC-free at age 70. To assess the impact of this decision, we repeated our analyses while removing these people. In an additional analysis, we again removed these people but used inverse probability of weighting to take them into account. The results did not change materially, although the estimated effect sizes and the standard errors were generally larger (Tables C.13 and C.14). 

To summarize our main findings, we have provided evidence for smoking effect on both  CRC MSI subtypes and evidence for heterogeneity under reasonable causal assumptions. Even though our case numbers were smaller and our follow-up was more limited than previous studies, the estimated $\sface$ lend support of a causal effect on non-MSI-high CRC, for which available evidence was limited.

\section{Discussion}
\label{Sec:Discuss}

Questions of etiologic heterogeneity are central to our understanding of how diseases evolve. Viewed through the lenses of competing events, existing  approaches largely overlooked formal discussions of causal estimands and identifiability assumptions. On the other hand, the recent progress in causal inference methodology for competing events presented causal estimands that either provide little information about etiology heterogeneity or estimands requiring identifiability assumptions that are implausible in the study of CRC subtypes.

In this paper, to study the effect of smoking on CRC subtypes defined by MSI status, we proposed the \sface s as causal effects that can complement existing methods and provide further insight towards the study of etiologic heterogeneity.   
One challenge with using multinomial regression with cohort data  is the issue of right censoring due to loss to follow-up or administrative censoring. To assess its impact,  we considered in this paper a number of analyses. The censoring problem would have been more pronounced if follow-up was longer than 10 years. Therefore, considering time-to-event outcomes is an important topic of future research. Another potential extension will consider continuous exposures instead of binary ones. Specifically, the generalization of the monotonicity assumptions, both in terms of definition and plausibility, is of interest.





\bibliographystyle{chicago}
\bibliography{SF_ACE}

\newcommand{\noop}[1]{}
\begin{thebibliography}{}

\bibitem[\protect\citeauthoryear{Amitay, Carr, Jansen, Roth, Alwers, Herpel,
  Kloor, Bl{\"a}ker, Chang-Claude, Brenner, and Hoffmeister}{Amitay
  et~al.}{2020}]{Amitay2020smoking}
Amitay, E.~L., P.~R. Carr, L.~Jansen, W.~Roth, E.~Alwers, E.~Herpel, M.~Kloor,
  H.~Bl{\"a}ker, J.~Chang-Claude, H.~Brenner, and M.~Hoffmeister (2020).
\newblock Smoking, alcohol consumption and colorectal cancer risk by molecular
  pathological subtypes and pathways.
\newblock {\em British Journal of Cancer\/}~{\em 122\/}(11), 1604--1610.

\bibitem[\protect\citeauthoryear{Bang and Robins}{Bang and
  Robins}{2005}]{bang2005doubly}
Bang, H. and J.~M. Robins (2005).
\newblock Doubly robust estimation in missing data and causal inference models.
\newblock {\em Biometrics\/}~{\em 61\/}(4), 962--973.

\bibitem[\protect\citeauthoryear{Begg, Seshan, and Zabor}{Begg
  et~al.}{2018}]{begg2018re}
Begg, C.~B., V.~E. Seshan, and E.~C. Zabor (2018).
\newblock Re:“a multinomial regression approach to model outcome
  heterogeneity”.
\newblock {\em American journal of epidemiology\/}~{\em 187\/}(5), 1129--1130.

\bibitem[\protect\citeauthoryear{Carr, Alwers, Bienert, Weberpals, Kloor,
  Brenner, and Hoffmeister}{Carr et~al.}{2018}]{carr2018lifestyle}
Carr, P., E.~Alwers, S.~Bienert, J.~Weberpals, M.~Kloor, H.~Brenner, and
  M.~Hoffmeister (2018).
\newblock Lifestyle factors and risk of sporadic colorectal cancer by
  microsatellite instability status: a systematic review and meta-analyses.
\newblock {\em Annals of Oncology\/}~{\em 29\/}(4), 825--834.

\bibitem[\protect\citeauthoryear{Chatterjee}{Chatterjee}{2004}]{chatterjee2004two}
Chatterjee, N. (2004).
\newblock A two-stage regression model for epidemiological studies with
  multivariate disease classification data.
\newblock {\em Journal of the American Statistical Association\/}~{\em
  99\/}(465), 127--138.

\bibitem[\protect\citeauthoryear{Frangakis and Rubin}{Frangakis and
  Rubin}{2002}]{frangakis2002principal}
Frangakis, C.~E. and D.~B. Rubin (2002).
\newblock Principal stratification in causal inference.
\newblock {\em Biometrics\/}~{\em 58\/}(1), 21--29.

\bibitem[\protect\citeauthoryear{Hern{\'a}n and Robins}{Hern{\'a}n and
  Robins}{2020}]{hernan2020causal}
Hern{\'a}n, M.~A. and J.~M. Robins (2020).
\newblock Causal inference: what if.
\newblock {\em Boca Raton: Chapman \& Hill/CRC\/}~{\em 2020}.

\bibitem[\protect\citeauthoryear{Liu, Nevo, Nishihara, Cao, Song, Twombly,
  Chan, Giovannucci, VanderWeele, Wang, and Ogino}{Liu
  et~al.}{2018}]{liu2018utility}
Liu, L., D.~Nevo, R.~Nishihara, Y.~Cao, M.~Song, T.~S. Twombly, A.~T. Chan,
  E.~L. Giovannucci, T.~J. VanderWeele, M.~Wang, and S.~Ogino (2018).
\newblock Utility of inverse probability weighting in molecular pathological
  epidemiology.
\newblock {\em European journal of epidemiology\/}~{\em 33\/}(4), 381--392.

\bibitem[\protect\citeauthoryear{Lunceford and Davidian}{Lunceford and
  Davidian}{2004}]{lunceford2004stratification}
Lunceford, J.~K. and M.~Davidian (2004).
\newblock Stratification and weighting via the propensity score in estimation
  of causal treatment effects: a comparative study.
\newblock {\em Statistics in medicine\/}~{\em 23\/}(19), 2937--2960.

\bibitem[\protect\citeauthoryear{Nevo, Nishihara, Ogino, and Wang}{Nevo
  et~al.}{2018}]{nevo2018competing}
Nevo, D., R.~Nishihara, S.~Ogino, and M.~Wang (2018).
\newblock The competing risks cox model with auxiliary case covariates under
  weaker missing-at-random cause of failure.
\newblock {\em Lifetime data analysis\/}~{\em 24\/}(3), 425--442.

\bibitem[\protect\citeauthoryear{Nevo, Ogino, and Wang}{Nevo
  et~al.}{2021}]{nevo2021reflection}
Nevo, D., S.~Ogino, and M.~Wang (2021).
\newblock Reflection on modern methods: causal inference considerations for
  heterogeneous disease etiology.
\newblock {\em International Journal of Epidemiology\/}.

\bibitem[\protect\citeauthoryear{Nevo, Zucker, Tamimi, and Wang}{Nevo
  et~al.}{2016}]{nevo2016accounting}
Nevo, D., D.~M. Zucker, R.~M. Tamimi, and M.~Wang (2016).
\newblock Accounting for measurement error in biomarker data and
  misclassification of subtypes in the analysis of tumor data.
\newblock {\em Statistics in medicine\/}~{\em 35\/}(30), 5686--5700.

\bibitem[\protect\citeauthoryear{Ogino, Nishihara, VanderWeele, Wang, Nishi,
  Lochhead, Qian, Zhang, Wu, Nan, et~al.}{Ogino et~al.}{2016}]{ogino2016role}
Ogino, S., R.~Nishihara, T.~J. VanderWeele, M.~Wang, A.~Nishi, P.~Lochhead,
  Z.~R. Qian, X.~Zhang, K.~Wu, H.~Nan, et~al. (2016).
\newblock The role of molecular pathological epidemiology in the study of
  neoplastic and non-neoplastic diseases in the era of precision medicine.
\newblock {\em Epidemiology (Cambridge, Mass.)\/}~{\em 27\/}(4), 602.

\bibitem[\protect\citeauthoryear{Robins}{Robins}{1986}]{robins1986new}
Robins, J. (1986).
\newblock A new approach to causal inference in mortality studies with a
  sustained exposure period—application to control of the healthy worker
  survivor effect.
\newblock {\em Mathematical modelling\/}~{\em 7\/}(9-12), 1393--1512.

\bibitem[\protect\citeauthoryear{Robins, Rotnitzky, and Zhao}{Robins
  et~al.}{1994}]{robins1994estimation}
Robins, J.~M., A.~Rotnitzky, and L.~P. Zhao (1994).
\newblock Estimation of regression coefficients when some regressors are not
  always observed.
\newblock {\em Journal of the American statistical Association\/}~{\em
  89\/}(427), 846--866.

\bibitem[\protect\citeauthoryear{Rubin}{Rubin}{2006}]{rubin2006causal}
Rubin, D.~B. (2006).
\newblock Causal inference through potential outcomes and principal
  stratification: application to studies with “censoring” due to death.
\newblock {\em Statistical Science\/}~{\em 21\/}(3), 299--309.

\bibitem[\protect\citeauthoryear{Stensrud, Young, Didelez, Robins, and
  Hern{\'a}n}{Stensrud et~al.}{2020}]{stensrud2020separable}
Stensrud, M.~J., J.~G. Young, V.~Didelez, J.~M. Robins, and M.~A. Hern{\'a}n
  (2020).
\newblock Separable effects for causal inference in the presence of competing
  events.
\newblock {\em Journal of the American Statistical Association\/}, 1--9.

\bibitem[\protect\citeauthoryear{Sun, VanderWeele, and Tchetgen~Tchetgen}{Sun
  et~al.}{2017}]{sun2017multinomial}
Sun, B., T.~VanderWeele, and E.~J. Tchetgen~Tchetgen (2017).
\newblock A multinomial regression approach to model outcome heterogeneity.
\newblock {\em American journal of epidemiology\/}~{\em 186\/}(9), 1097--1103.

\bibitem[\protect\citeauthoryear{Sun, VanderWeele, and Tchetgen~Tchetgen}{Sun
  et~al.}{2018}]{sun2018reply}
Sun, B., T.~VanderWeele, and E.~J. Tchetgen~Tchetgen (2018, 03).
\newblock {THE AUTHORS REPLY}.
\newblock {\em American Journal of Epidemiology\/}~{\em 187\/}(5), 1130--1131.

\bibitem[\protect\citeauthoryear{Ugai, V{\"a}yrynen, Haruki, Akimoto, Lau,
  Zhong, Kishikawa, V{\"a}yrynen, Zhao, Fujiyoshi, et~al.}{Ugai
  et~al.}{2022}]{ugai2022smoking}
Ugai, T., J.~P. V{\"a}yrynen, K.~Haruki, N.~Akimoto, M.~C. Lau, R.~Zhong,
  J.~Kishikawa, S.~A. V{\"a}yrynen, M.~Zhao, K.~Fujiyoshi, et~al. (2022).
\newblock Smoking and incidence of colorectal cancer subclassified by
  tumor-associated macrophage infiltrates.
\newblock {\em JNCI: Journal of the National Cancer Institute\/}~{\em
  114\/}(1), 68--77.

\bibitem[\protect\citeauthoryear{Wang, Kuchiba, and Ogino}{Wang
  et~al.}{2015}]{wang2015meta}
Wang, M., A.~Kuchiba, and S.~Ogino (2015).
\newblock A meta-regression method for studying etiological heterogeneity
  across disease subtypes classified by multiple biomarkers.
\newblock {\em American journal of epidemiology\/}~{\em 182\/}(3), 263--270.

\bibitem[\protect\citeauthoryear{Wang, Spiegelman, Kuchiba, Lochhead, Kim,
  Chan, Poole, Tamimi, Tworoger, Giovannucci, et~al.}{Wang
  et~al.}{2016}]{wang2016statistical}
Wang, M., D.~Spiegelman, A.~Kuchiba, P.~Lochhead, S.~Kim, A.~T. Chan, E.~M.
  Poole, R.~Tamimi, S.~S. Tworoger, E.~Giovannucci, et~al. (2016).
\newblock Statistical methods for studying disease subtype heterogeneity.
\newblock {\em Statistics in medicine\/}~{\em 35\/}(5), 782--800.

\bibitem[\protect\citeauthoryear{Young, Stensrud, Tchetgen~Tchetgen, and
  Hern{\'a}n}{Young et~al.}{2020}]{young2020causal}
Young, J.~G., M.~J. Stensrud, E.~J. Tchetgen~Tchetgen, and M.~A. Hern{\'a}n
  (2020).
\newblock A causal framework for classical statistical estimands in
  failure-time settings with competing events.
\newblock {\em Statistics in Medicine\/}~{\em 39\/}(8), 1199--1236.

\bibitem[\protect\citeauthoryear{Zabor and Begg}{Zabor and
  Begg}{2017}]{zabor2017comparison}
Zabor, E.~C. and C.~B. Begg (2017).
\newblock A comparison of statistical methods for the study of etiologic
  heterogeneity.
\newblock {\em Statistics in medicine\/}~{\em 36\/}(25), 4050--4060.

\bibitem[\protect\citeauthoryear{Zehavi and Nevo}{Zehavi and
  Nevo}{2021}]{zehavi2021matching}
Zehavi, T. and D.~Nevo (2021).
\newblock A matching framework for truncation by death problems.
\newblock {\em arXiv preprint arXiv:2110.10186\/}.

\bibitem[\protect\citeauthoryear{Zhang and Rubin}{Zhang and
  Rubin}{2003}]{zhang2003estimation}
Zhang, J.~L. and D.~B. Rubin (2003).
\newblock Estimation of causal effects via principal stratification when some
  outcomes are truncated by “death”.
\newblock {\em Journal of Educational and Behavioral Statistics\/}~{\em
  28\/}(4), 353--368.

\end{thebibliography}
\appendix









\renewcommand\thefigure{\thesection.\arabic{figure}}    
\renewcommand{\thetable}{\Alph{section}.\arabic{table}}
\renewcommand\theequation{A.\arabic{equation}}


\section*{Appendix}
\begin{itemize}
\item Section \ref{AppSec:Theory}  includes proofs and additional theory. Section \ref{AppSubSec:PropProofs} presents the proofs of Propositions 1--3. Section \ref{AppSubSec:RelaxMono} illustrates the the connection between observed data and the potential outcomes. Section \ref{AppSubSec:AddSens} discuss identification results under weaker combination of assumptions. Section \ref{AppSubSec:DR} provides the complete expression for the DR estimator. 
\item Section B includes details on the  simulations and presents the results. Section \ref{AppSec:sim_dgm} gives additional details about the data generating mechanism of the simulation. Section \ref{AppSec:sim_studies} presents the details of  three different simulation studies. Finally, Section \ref{AppSec:sim_res}  presents tables and figures summarizing the different simulation results.
\item Section C presents details on the data  as well as additional analyses and results. Section \ref{AppSec:data_info} includes descriptive information about the data used. Section \ref{AppSec:data_models_info} gives additional information about the weights used for handling missing subtypes and about the models fitted for estimating effects. Finally, Section \ref{AppSec:data_res_info} presents additional  data analysis results. 
\end{itemize}
\appendix

\section{Proofs and additional theory}
\label{AppSec:Theory}

\subsection{Proof of identification results} 
\label{AppSubSec:PropProofs}

Before providing the identification formulas for the $\sface$, we first show that the $\sface$ equals to the $TE$ when defined on the RR scale under \Smono for both disease subtypes.

\subsubsection{Equivalence of the $\sface$ and TE}

When defined on the RR scale, the $\sface^{(1)}$ equals to
\begin{align}
\sface^{(1)}_{RR} = & \frac{\bbE  \left[ Y^{(1)}(1)\big|\  Y^{(2)}(0)=Y^{(2)}(1)=0 \right]}{\bbE \left[Y^{(1)}(0) \big| \ Y^{(2)}(0)=Y^{(2)}(1)=0 \right]} \notag\\[1em] 
= & \frac{\bbE  \left[ Y^{(1)}(1)\big|\  Y^{(2)}(1)=0 \right]}{\bbE \left[Y^{(1)}(0) \big| \ Y^{(2)}(1)=0 \right]} \label{prop1RR:line-2}\\[1em] 
= & \frac{\Pr[Y^{(1)}(1) = 1, Y^{(2)}(1) = 0 ]}{\Pr[Y^{(1)}(0) = 1, Y^{(2)}(1) = 0]} \notag\\[1em]  
= & \frac{\Pr[Y^{(1)}(1) = 1]}{\Pr[Y^{(1)}(0) = 1]}. \label{prop1RR:line-4}
\end{align}
Equality \eqref{prop1RR:line-2} holds by the \Smono assumption for subtype 2. Equality \eqref{prop1RR:line-4} holds because the subtypes are mutually exclusive (numerator) and by the \Smono assumption for subtype 1  (denominator).

\subsubsection{Proof of Proposition 1} 
Starting from the definition of $\sface$, we have 
\begin{align}
\sface^{(1)}_D = & \bbE  \big[ Y^{(1)}(1) - Y^{(1)}(0) \big| Y^{(2)}(0)=Y^{(2)}(1)=0 \big] \notag\\[1em] 
= & \mathbb{E}  \left[ Y^{(1)}(1) - Y^{(1)}(0) \big| Y^{(2)}(1)=0 \right] \label{prop1:line-2}\\[1em] 
= & \frac{\Pr[Y^{(1)}(1) = 1, Y^{(2)}(1) = 0 ] - \Pr[Y^{(1)}(0) = 1, Y^{(2)}(1) = 0]}{\Pr[Y^{(2)}(1) = 0]} \notag\\[1em]  
= & \frac{\Pr[Y^{(1)}(1) = 1] - \Pr[Y^{(1)}(0) = 1]}{\Pr[Y^{(2)}(1) = 0]},  \label{prop1:line-4}
\end{align}
Equality \eqref{prop1:line-2} follows from the \Smono assumption for subtype 2, and \eqref{prop1:line-4} is justified because the subtypes are mutually exclusive (first expression in the numerator) and by \Smono for subtype 1 (second expression in the numerator). 

For the standardization-based identification formula, we continue by recalling that under positivity and weak ignorability,  $\Pr[Y^{(k)}(a) = j]$ is identified by 
$$
\Pr[Y^{(k)}(a) = j]=\bbE_{\bX}[\Pr(Y^{(k)}=j|A=a,\bX)]
$$ 
for $a,j,k=0,1$ \citep{hernan2020causal}. With $\Pr(Y^{(k)}=j|A=a,\bx)$ previously denoted by $\pi_k(a,\bx)$ for $j=1$ and hence equals to $1-\pi_k(a,\bx)$ for $j=0$, we substitute these expressions in \eqref{prop1:line-4} to get
\begin{equation}
\sface_D = \frac{\bbE_{\bX}[ \pi_1(1,\bX)] - \bbE_{\bX}[\pi_1(0,\bX)]}{1-\bbE_{\bX}
[\pi_2(1,\bX)]}. \label{prop1stand:line-2}
\end{equation}
Alternatively, to get IPTW-based identification results, we may continue from \eqref{prop1:line-4} by recalling that under positivity and weak ignorability,  $\Pr[Y^{(k)}(a) = 1]$ is identified by $\Pr[Y^{(k)}(a) = 1]=\mathbb{E}\left[ \frac{ \mathbbm{1} \{A=a\} Y^{(k)}}{\Pr(A=a|\bX)}\right] $ \citep{hernan2020causal}. Substituting these expressions in \eqref{prop1:line-4}, we get 
\begin{equation}
\sface^{(1)}_D =\frac{\bbE\left[ \frac{\bbmI \{A=1\} Y^{(1)}}{e(\bX)}\right] - \bbE\left[ \frac{\bbmI \{A=0\} Y^{(1)}}{1-e(\bX)}\right]}{1 - \bbE\left[ \frac{\bbmI \{A=1\} Y^{(2)}}{e(\bX)}\right]}. \label{propo1iptw:line-6}
\end{equation}
The identification formulas for subtype 2 are analogous. Using standardization,
\begin{equation*}
\sface^{(2)}_D= \frac{\bbE_{\bX} [\pi_2(1,\bX)] - \bbE_{\bX} [\pi_2(0,\bX)]}{1- \bbE_{\bX}[ \pi_1(0,\bX)]}, 
\end{equation*}
or using IPTW, 
\begin{equation*}
\sface^{(2)}_D
 =\frac{\bbE\left[ \frac{\bbmI \{A=1\} Y^{(2)}}{e(\bX)}\right] - \bbE\left[ \frac{\bbmI \{A=0\} Y^{(2)}}{1-e(\bX)}\right]}{1 - \bbE\left[ \frac{\bbmI \{A=1\} Y^{(1)}}{e(\bX)}\right]}. 
\end{equation*}
The difference between the effects is given by substituting the expressions above in
$\theta_D = \sface^{(1)}_D -\sface^{(2)}_D$

The RR-based measures \eqref{prop1RR:line-4} are identified under weak ignorability and positivity  using standardization by
\begin{equation*}
\sface^{(k)}_{RR}=   \frac{\bbE_{\bX} [\pi_k(1,\bX)] }{\bbE_{\bX} [\pi_k(0,\bX)]} 
\end{equation*}
for $k=1,2$, or using IPTW, 
\begin{equation*}
\sface^{(k)}_{RR}
=\frac{\bbE\left[ \frac{ \bbmI \{A=1\} Y^{(k)}}{e(\bX)}\right] }{ \bbE\left[ \frac{ \bbmI \{A=0\} Y^{(k)}}{1-e(\bX)}\right]}.
\end{equation*}
Finally, define $\theta_{RR}$ to be the difference between the effects in the RR scale. So $\theta_{RR}$ is given by
\begin{equation*}
\theta_{RR} = \sface^{(1)}_{RR} -\sface^{(2)}_{RR}.
\end{equation*}

\subsubsection{Proof of Proposition 2}
Recall that  $\lambda_1 = \Pr[Y^{(2)}(1) = 1| \ Y^{(1)}(0) = 1]$ and $\lambda_2 = \Pr[Y^{(1)}(1) = 1| \ Y^{(2)}(0) = 1]$. 
We start with the effect on subtype 1. \begin{align}
&\sface^{(1)}_D \notag\\
&= \bbE[Y^{(1)}(1)- Y^{(1)}(0) | Y^{(2)}(0) = Y^{(2)}(1) = 0] \notag\\[1em]
& = \frac{\Pr[Y^{(1)}(1) = 1,  Y^{(2)}(0) = 0, Y^{(2)}(1) = 0] - \Pr[Y^{(1)}(0) = 1,  Y^{(2)}(0) = 0, Y^{(2)}(1) = 0]}{\Pr[Y^{(2)}(0) = Y^{(2)}(1) = 0]} \notag\\[1em]   
& = \frac{\Pr[Y^{(1)}(1) = 1,  Y^{(2)}(0) = 0] - \Pr[Y^{(1)}(0) = 1, Y^{(2)}(1) = 0]}{\Pr[Y^{(2)}(1) = 0] - \Pr[Y^{(2)}(0) = 1, Y^{(2)}(1) = 0]} \label{prop2:line-3}\\[1em] 
& = \frac{\Pr[Y^{(1)}(1) = 1] - \Pr[Y^{(1)}(1) = 1,  Y^{(2)}(0) = 1] - \Pr[Y^{(1)}(0) = 1]+\Pr[Y^{(1)}(0) = 1, Y^{(2)}(1) = 1]}{\Pr[Y^{(2)}(1) = 0] - \Pr[Y^{(2)}(0) = 1,  Y^{(1)}(1) = 1]} \label{prop2:line-4}\\[1em]
& = \frac{\Pr[Y^{(1)}(1) = 1] - \lambda_2 \Pr[Y^{(2)}(0) = 1] - \Pr[Y^{(1)}(0) = 1]+ \lambda_1 \Pr[Y^{(1)}(0) = 1]}{\Pr[Y^{(2)}(1) = 0] - \lambda_2 \Pr[Y^{(2)}(0) = 1]}  \notag\\[1em]
& = \frac{\Pr[Y^{(1)}(1) = 1] + (\lambda_1 - 1) \Pr[Y^{(1)}(0) = 1] - \lambda_2 \Pr[Y^{(2)}(0) = 1]}{\Pr[Y^{(2)}(1) = 0] - \lambda_2 \Pr[Y^{(2)}(0) = 1]}.  \label{prop2:line-6}
\end{align}
In \eqref{prop2:line-3}, the numerator follows from the fact the subtypes are mutually exclusive and the denominator is justified by the law of total probability. In \eqref{prop2:line-4}, the numerator holds because of the law of total probability and the denominator follows by noting that the \Dmono assumption for subtype 2 implies $\Pr[Y^{(2)}(0) = 1, Y^{(2)}(1) = 0] = \Pr[Y^{(2)}(0) = 1,  Y^{(1)}(1) = 1]$.

Now, similarly to the proof of Proposition 1, we can continue using standardization,
\begin{equation*}
\sface^{(1)}_D=  \frac{\bbE_{\bX}[\pi_1(1,\bX)] +(\lambda_1 - 1) \bbE_{\bX}[\pi_1(0,\bX)] - \lambda_2  \bbE_{\bX}[\pi_2(0,\bX)]} { 1-\bbE_{\bX}[\pi_2(1,\bX)] - \lambda_2  \bbE_{\bX}[\pi_2(0,\bX)]} 
\end{equation*}
by weak ignorability and positivity. Alternatively, as in the proof of Proposition 1, we can continue from \eqref{prop2:line-6} using IPTW to obtain
\begin{equation*}
    \sface_D^{(1)}
 =\frac{\bbE\left[ \frac{\bbmI \{A=1\} Y^{(1)}}{e(\bX)}\right] + (\lambda_1-1) \bbE\left[ \frac{ \bbmI \{A=0\} Y^{(1)}}{1-e(\bX)}\right]-\lambda_2\bbE\left[\frac{\mathbbm{1}\{A=0\}Y^{(2)}}{1-e(\bX)}\right]}{1 - \bbE\left[\frac{\bbmI \{A=1\} Y^{(2)}}{e(\bX)}\right]-\lambda_2\bbE\left[\frac{\bbmI\{A=0\}Y^{(2)}}{1-e(\bX)}\right]}, \label{prop2iptw:line-2}
\end{equation*}
again by weak ignorability and positivity, similarly to the standard IPTW-based identification.

The identification formulas for subtype 2 are analogous. Using standardization, 
\begin{equation*}
\sface^{(2)}_D= \frac{\bbE_{\bX}[\pi_2(1,\bX)] +(\lambda_1 - 1) \bbE_{\bX}[\pi_2(0,\bX)] - \lambda_2  \bbE_{\bX}[\pi_1(0,\bX)]} { 1-\bbE_{\bX}[\pi_1(1,\bX)] - \lambda_2  \bbE_{\bX}[\pi_1(0,\bX)]} 
\end{equation*}
or using IPTW,
\begin{equation*}
\sface^{(2)}_D =\frac{\bbE\left[ \frac{\bbmI \{A=1\} Y^{(2)}}{e(\bX)}\right] + (\lambda_1-1) \bbE\left[ \frac{ \bbmI \{A=0\} Y^{(2)}}{1-e(\bX)}\right]-\lambda_2\bbE\left[\frac{\bbmI\{A=0\}Y^{(1)}}{1-e(\bX)}\right]}{1 - \bbE\left[\frac{\bbmI \{A=1\} Y^{(1)}}{e(\bX)}\right]-\lambda_2\bbE\left[\frac{\bbmI\{A=0\}Y^{(1)}}{1-e(\bX)}\right]}. 
\end{equation*}
The difference between the effects is given by
\begin{equation*}
\theta_D = \sface^{(1)}_D - \sface^{(2)}_D.
\end{equation*}

\subsubsection{Proof of Proposition 3}
Recall that  $\lambda_1 = \Pr[Y^{(2)}(1) = 1| \ Y^{(1)}(0) = 1]$ and $\lambda_2 = \Pr[Y^{(1)}(1) = 1| \ Y^{(2)}(0) = 1]$. 
We start with the effect on subtype 1.
\begin{align}
\sface^{(1)}_{RR} & =  \frac{\bbE  \left[ Y^{(1)}(1)\big| Y^{(2)}(0)=Y^{(2)}(1)=0 \right]}{\bbE \left[Y^{(1)}(0) \big| Y^{(2)}(0)=Y^{(2)}(1)=0 \right]} \notag\\[1em] 
& = \frac{\Pr[Y^{(1)}(1) = 1,  Y^{(2)}(0) = 0, Y^{(2)}(1) = 0]}{\Pr[Y^{(1)}(0) = 1,  Y^{(2)}(0) = 0, Y^{(2)}(1) = 0]} \notag\\[1em]   
& = \frac{\Pr[Y^{(1)}(1) = 1,  Y^{(2)}(0) = 0]}{\Pr[Y^{(1)}(0) = 1, Y^{(2)}(1) = 0]} \label{prop2RR:line-2}\\[1em] 
& = \frac{\Pr[Y^{(1)}(1) = 1] - \Pr[Y^{(1)}(1) = 1,  Y^{(2)}(0) = 1]}{ \Pr[Y^{(1)}(0) = 1]-\Pr[Y^{(1)}(0) = 1, Y^{(2)}(1) = 1]} \label{prop2RR:line-3}\\[1em]
& = \frac{\Pr[Y^{(1)}(1) = 1] - \lambda_2 \Pr[Y^{(2)}(0) = 1]}{\Pr[Y^{(1)}(0) = 1]- \lambda_1 \Pr[Y^{(1)}(0) = 1]} \notag\\[1em]
& = \frac{\Pr[Y^{(1)}(1) = 1] - \lambda_2 \Pr[Y^{(2)}(0) = 1] }{(1-\lambda_1) \Pr[Y^{(1)}(0) = 1]}. \notag
\end{align}
Equality \eqref{prop2RR:line-2} holds because the subtypes are mutually exclusive. Equality \eqref{prop2RR:line-3} is by the law of total probability.
As before, using standardization, we may write
\begin{align}
\sface^{(1)}_{RR}=  & \frac{\bbE_{\bX}[\pi_1(1,\bX)] - \lambda_2  \bbE_{\bX}[\pi_2(0,\bX)]} {(1-\lambda_1) \bbE_{\bX}[\pi_1(0,\bX)] }\notag
\end{align}
or using IPTW, 
\begin{equation*}
\sface^{(1)}_{RR}
 =\frac{\bbE\left[ \frac{\bbmI \{A=1\} Y^{(1)}}{e(\bX)}\right]-\lambda_2\bbE\left[\frac{\mathbbm{1}\{A=0\}Y^{(2)}}{1-e(\bX)}\right]}{(1-\lambda_1) \bbE\left[ \frac{ \bbmI \{A=0\} Y^{(1)}}{1-e(\bX)}\right]}. \notag
\end{equation*}
The identification formula of subtype 2 is analogous. Using standardization, 
\begin{equation*}
\sface^{(2)}_{RR}=   \frac{\bbE_{\bX}[\pi_2(1,\bX)] - \lambda_1  \bbE_{\bX}[\pi_1(0,\bX)]} {(1-\lambda_2) \bbE_{\bX}[\pi_2(0,\bX)] },
\end{equation*}
or using IPTW, 
\begin{equation*}
\sface^{(2)}_{RR}
 =\frac{\bbE\left[ \frac{\bbmI \{A=1\} Y^{(2)}}{e(\bX)}\right]-\lambda_1\bbE\left[\frac{\mathbbm{1}\{A=0\}Y^{(1)}}{1-e(\bX)}\right]}{(1-\lambda_2) \bbE\left[ \frac{ \bbmI \{A=0\} Y^{(2)}}{1-e(\bX)}\right]}\notag
\end{equation*}
The difference between the effects is given by $\theta_{RR} = \sface^{(1)}_{RR} -\sface^{(2)}_{RR}$.

An interesting distinction between the results in Propositions 2 and 3 compared to Proposition 1, is that if one further assumes \Smono for one of the subtypes only, such that \Smono holds for one subtype only, say subtype 1 so $\lambda_1=0$, then the identification formulas for the two subtypes are no longer symmetric.

\subsection{Relationship between observed data and potential outcome profiles}
\label{AppSubSec:RelaxMono}

Table 1 of the main text defines the nine possible potential outcome profiles. Table \ref{Tab:obs2po} specifies the possible profiles for each participant according to their observed data $(A, Y^{(1)}, Y^{(2)})$  under each of the four combinations of assumptions. 

\begin{table}[ht]
\captionsetup{font=footnotesize}
\centering
\small
\addvbuffer[12pt] {\begin{tabular}{ P{2.5cm}  P{2.5cm} P{2.5cm} P{2.5cm} P{2.5cm} }
\hline
Observed data & \multicolumn{4}{c}{Monotonicity assumptions} \\
\hline
$(A, Y^{(1)}, Y^{(2)})$ & \thead{$Y^{(1)}$ S-Mono\\$Y^{(2)}$ S-Mono} & \thead{$Y^{(1)}$ D-Mono\\$Y^{(2)}$ D-Mono} & \thead{$Y^{(1)}$ S-Mono\\$Y^{(2)}$ D-Mono}  & 
\thead{$Y^{(1)}$ D-Mono\\$Y^{(2)}$ S-Mono} \\
\hline
$(0, 0, 0)$  & 0,1,3 & 0,1,3 & 0,1,3 & 0,1,3 \\ 
$(0, 1, 0)$  & 5 & 5,7 & 5 & 5,7 \\
$(0, 0, 1)$  & 6 & 6,8 & 6,8 & 6 \\
$(1, 0, 0)$ & 0 & 0 & 0 & 0  \\
$(1, 1, 0)$ & 3, 5 &  3, 5,8 & 3,5,8 & 3, 5 \\
$(1, 0, 1)$ & 1, 6 & 1,6,7 & 1,6 & 1,6,7 \\
\hline
\end{tabular}}
\caption{Connection between observed data and possible potential outcome profiles defined in Table 1 under four combinations of monotonicity assumptions.}
\label{Tab:obs2po}
\end{table}

\subsection{Additional identification results}
\label{AppSubSec:AddSens}
\subsubsection{\Smono for one subtype and no monotonicity assumption for the other subtype}
We present here identification for the $\sface_D^{(k)}$ under \Smono for subtype 1 and no monotonicity assumption, of any version, for subtype 2. To this end, in addition to $\lambda_2$, we define an additional sensitivity parameter $\lambda_2^0 = \Pr(Y^{(2)}(1) = 0, Y^{(1)}(1) = 0| Y^{(2)}(0) = 1)$. This sensitivity parameter is the probability to be CRC-free when exposed among those who would have been diagnosed with subtype 2 had they were unexposed. Under either \Smono or \DmonoNO, $\lambda_2^0=0$.

We begin with the effect on subtype 1,
\begin{align}
& \sface^{(1)}_D  = \bbE[Y^{(1)}(1)- Y^{(1)}(0) | Y^{(2)}(0) = Y^{(2)}(1) = 0] \notag\\[1em]
& = \frac{\Pr[Y^{(1)}(1) = 1,  Y^{(2)}(0) = 0, Y^{(2)}(1) = 0] - \Pr[Y^{(1)}(0) = 1,  Y^{(2)}(0) = 0, Y^{(2)}(1) = 0]}{\Pr[Y^{(2)}(0) = Y^{(2)}(1) = 0]}  \notag\\[1em]  
& = \frac{\Pr[Y^{(1)}(1) = 1,  Y^{(2)}(0) = 0] - \Pr[Y^{(1)}(0) = 1]}{\Pr[Y^{(2)}(0) = Y^{(2)}(1) = 0]}  \label{add_sens_s1:line-3}\\[1em]          
& = \frac{\Pr[Y^{(1)}(1) = 1] - \Pr[Y^{(1)}(1) = 1,  Y^{(2)}(0) = 1] - \Pr[Y^{(1)}(0) = 1]}{\Pr[Y^{(2)}(1) = 0] - \Pr[Y^{(2)}(1) = 0, Y^{(2)}(0) = 1]}  \label{add_sens_s1:line-4}\\[1em]   
& = \frac{\Pr[Y^{(1)}(1) = 1] - \Pr[Y^{(1)}(1) = 1,  Y^{(2)}(0) = 1] - \Pr[Y^{(1)}(0) = 1]}{\Pr[Y^{(2)}(1) = 0] - \Pr[Y^{(2)}(0) = 1, Y^{(1)}(1) = 1] - \Pr[Y^{(2)}(1) = 0, Y^{(1)}(1) = 0, Y^{(2)}(0) = 1]}  \label{add_sens_s1:line-5}\\[1em]
& = \frac{\Pr[Y^{(1)}(1) = 1] - \lambda_2 \Pr[Y^{(2)}(0) = 1] - \Pr[Y^{(1)}(0) = 1]}{\Pr[Y^{(2)}(1) = 0] - \lambda_2 \Pr[Y^{(2)}(0) = 1] - \lambda_2^0 \Pr[Y^{(2)}(0) = 1]}  \notag
\end{align}
Equality \eqref{add_sens_s1:line-3} holds because the subtypes are mutually exclusive and because of the equality $\Pr[Y^{(1)}(0) = 1,  Y^{(2)}(0) = 0, Y^{(2)}(1) = 0] = \Pr[Y^{(1)}(0) = 1]$ which holds due to the \Smono assumption for subtype 1. Equality \eqref{add_sens_s1:line-4} holds because of the law of total probability. Equality \eqref{add_sens_s1:line-5} holds because of the law of total probability and because the subtypes are mutually exclusive. 

Because different assumptions were made for subtype 1 and subtype 2, the effects are not analogous. Turning to the effect on subtype 2, 
\begin{align}
& \sface^{(2)}_D  = \bbE[Y^{(2)}(1)- Y^{(2)}(0) | Y^{(1)}(0) = Y^{(1)}(1) = 0] \notag\\[1em]
&   = \bbE[Y^{(2)}(1)- Y^{(2)}(0) | Y^{(1)}(1) = 0] \label{add_sens_s2:line-2}\\[1em]
& = \frac{\Pr[Y^{(2)}(1) = 1, Y^{(1)}(1) = 0] - \Pr[Y^{(2)}(0) = 1, Y^{(1)}(1) = 0]}{\Pr[Y^{(1)}(1) = 0]}  \notag\\[1em]  
& = \frac{\Pr[Y^{(2)}(1) = 1] - \Pr[Y^{(2)}(0) = 1, Y^{(1)}(1) = 0]}{\Pr[Y^{(1)}(1) = 0]}   \label{add_sens_s2:line-4}\\[1em]  
& = \frac{\Pr[Y^{(2)}(1) = 1] - \Pr[Y^{(2)}(0) = 1] + \Pr[Y^{(2)}(0) = 1, Y^{(1)}(1) = 1]}{\Pr[Y^{(1)}(1) = 0]}  \label{add_sens_s2:line-5}\\[1em]  
& = \frac{\Pr[Y^{(2)}(1) = 1] - \Pr[Y^{(2)}(0) = 1] + \lambda_2 \Pr[Y^{(2)}(0) = 1]}{\Pr[Y^{(1)}(1) = 0]}  \notag
\end{align}
Equality \eqref{add_sens_s2:line-2} holds because of the \Smono assumption for subtype 1. Equality \eqref{add_sens_s2:line-4} holds because the subtypes are mutually exclusive. Equality \eqref{add_sens_s2:line-5} holds because of the law of total probability. As before, standardization, IPTW or DR estimation can be used to identify and estimate the effects.  

 
\subsubsection{Sensitivity analysis without monotonicity assumptions}
We present here identification on the difference scale, for the $\sface^{(k)}$ under no specified monotonicity assumptions for both subtypes. Recall that  $\lambda_1 = \Pr[Y^{(2)}(1) = 1| \ Y^{(1)}(0) = 1]$, $\lambda_2 = \Pr[Y^{(1)}(1) = 1| \ Y^{(2)}(0) = 1]$ and $\lambda_2^0 = \Pr(Y^{(2)}(1) = 0, Y^{(1)}(1) = 0| Y^{(2)}(0) = 1)$. 

We begin with the effect on subtype 1,
\begin{align}
& \sface^{(1)}_D  = \bbE[Y^{(1)}(1)- Y^{(1)}(0) | Y^{(2)}(0) = Y^{(2)}(1) = 0] \notag\\[1em]
& = \frac{\Pr[Y^{(1)}(1) = 1,  Y^{(2)}(0) = 0, Y^{(2)}(1) = 0] - \Pr[Y^{(1)}(0) = 1,  Y^{(2)}(0) = 0, Y^{(2)}(1) = 0]}{\Pr[Y^{(2)}(0) = Y^{(2)}(1) = 0]}  \notag\\[1em]  
& = \frac{\Pr[Y^{(1)}(1) = 1,  Y^{(2)}(0) = 0] - \Pr[Y^{(1)}(0) = 1, Y^{(2)}(1) = 0]}{\Pr[Y^{(2)}(0) = Y^{(2)}(1) = 0]}  \label{add_sens_no1:line-3}\\[1em]          
& = \frac{\Pr[Y^{(1)}(1) = 1] - \Pr[Y^{(1)}(1) = 1,  Y^{(2)}(0) = 1] - \Pr[Y^{(1)}(0) = 1, Y^{(2)}(1) = 0]}{\Pr[Y^{(2)}(1) = 0] - \Pr[Y^{(2)}(1) = 0, Y^{(2)}(0) = 1]}  \label{add_sens_no1:line-4}\\[1em]   
& = \frac{\Pr[Y^{(1)}(1) = 1] - \Pr[Y^{(1)}(1) = 1,  Y^{(2)}(0) = 1] - \Pr[Y^{(1)}(0) = 1] + \Pr[Y^{(1)}(0) = 1, Y^{(2)}(1) = 1]}{\Pr[Y^{(2)}(1) = 0] - \Pr[Y^{(2)}(0) = 1, Y^{(1)}(1) = 1] - \Pr[Y^{(2)}(1) = 0, Y^{(1)}(1) = 0, Y^{(2)}(0) = 1]}  \label{add_sens_no1:line-5}\\[1em]
& = \frac{\Pr[Y^{(1)}(1) = 1] - \lambda_2 \Pr[Y^{(2)}(0) = 1] -(1-\lambda_1) \Pr[Y^{(1)}(0) = 1]}{\Pr[Y^{(2)}(1) = 0] - \lambda_2 \Pr[Y^{(2)}(0) = 1] - \lambda_2^0 \Pr[Y^{(2)}(0) = 1]}  \notag
\end{align}

Equality \eqref{add_sens_no1:line-3} holds because the subtypes are mutually exclusive. Equality \eqref{add_sens_no1:line-4} holds because of the law of total probability. Equality \eqref{add_sens_no1:line-5} holds because of the law of total probability and because the subtypes are mutually exclusive. 

The identification formula of subtype 2 is analogous, and requires one additional parameter $\lambda_1^0 = \Pr(Y^{(1)}(1) = 0, Y^{(2)}(1) = 0| Y^{(1)}(0) = 1)$.
\begin{align}
& \sface^{(2)}_D  = \frac{\Pr[Y^{(2)}(1) = 1] - \lambda_1 \Pr[Y^{(1)}(0) = 1] -(1-\lambda_2) \Pr[Y^{(2)}(0) = 1]}{\Pr[Y^{(1)}(1) = 0] - \lambda_1 \Pr[Y^{(1)}(0) = 1] - \lambda_1^0 \Pr[Y^{(1)}(0) = 1]}.  \notag
\end{align}
 As before, standardization, IPTW or DR estimation can be used to identify and estimate the effects.

\subsection{Details about the doubly-robust estimator}
\label{AppSubSec:DR}

The doubly-robust estimators that are plugged in Equations \eqref{prop1:line-4} (under \Smono for both subtypes), \eqref{prop2:line-6} (under \Dmono for both subtypes), or their RR analogues are given by, 
\begin{align*}
\widehat{\Pr}[Y^{(k)}(1)=1] &= \frac{1}{n}\sum_{i=1}^n\left[\frac{A_iY_i^{(k)}}{e(\bX_i;\hat{\eta}_2)} - \frac{[A_i-e(\bX_i;\hat{\eta}_2)]\frac{1}{n}\sum_{i=1}^n\left[\pi_k(1,\bX_i;\hat{\eta}_1)\right]}{e(\bX_i;\hat{\eta}_2)} \right] \\
\widehat{\Pr}[Y^{(k)}(0)=1] &= \frac{1}{n}\sum_{i=1}^n\left[\frac{(1-A_i)Y_i^{(k)}}{1-e(\bX_i;\hat{\eta}_2)} + \frac{[A_i-e(\bX_i;\hat{\eta}_2)]\frac{1}{n}\sum_{i=1}^n\left[\pi_k(1,\bX_i;\hat{\eta}_1)\right]}{1-e(\bX_i;\hat{\eta}_2)} \right].
\end{align*}



\newpage
\section{Additional simulation details and results}
\label{AppSec:AppSims}
Section \ref{AppSec:sim_dgm} gives additional details about the data generating mechanism of the simulation. Section \ref{AppSec:sim_studies} presents the details of  three different simulation studies. Finally, Section \ref{AppSec:sim_res}  presents tables and figures summarizing the different simulation results.

\subsection{Data generating mechanism}
\label{AppSec:sim_dgm}

The data were simulated to resemble an observational study. For each individual $i$, we first simulated two independent measured confounders, $\bX_i=(X_{i1},X_{i2})$, where $X_1 \sim N(0,1)$ and $X_2 \sim Ber(0.5)$ and one normally distributed unmeasured covariate $U_i \sim N(0,1)$.  To simulate the potential outcomes, we used the multinomial regression model
\begin{equation}
\label{sim_multi_model}
\Pr(Y_i(a)=k|\bX_i=\bx, U_i=u) = \frac{\exp(\alpha_k + \beta_k a + \bgamma_k^T \bx + \delta_k u)}{1 + \sum_{j=1}^{2} \exp(\alpha_j + \beta_j a + \bgamma_j^T  \bx  + \delta_k u)},
\end{equation}
for $k=1,2$, and 
\small
$$
\Pr(Y_i(a)=0|\bX_i=\bx, U_i=u)  = 1 
- \sum_{k=1}^{2}\Pr(Y_i(a)=k|\bX_i=\bx, U_i=u). 
$$
We first simulated the potential outcome under no exposure $Y_i(0)$, for each individual $i$, using model \eqref{sim_multi_model}. To make sure the data followed the \Smono assumption for both disease subtypes, the potential outcomes under exposure ($a=1$) were created as follows: Individuals with $Y_i(0)>0$, namely that they would have been diagnosed with disease subtype 1 or 2 under $a=0$, would automatically be diagnosed with the same disease subtype under $a=1$. For individuals who were disease-free under $a=0$ the disease subtype under $a=1$ was simulated using Equation \eqref{sim_multi_model} with an adjusted probability using the Law of Total Probability,
$$
\Pr[Y_i(1)=k|\bX_i=\bx, U_i = u, Y_i(0)=0] = \frac{\Pr[Y_i(1)=k|\bX_i=\bx, U=u) - \Pr(Y_i(0)=k|\bX_i=\bx, U=u]}{\Pr[Y_i(0)=0|\bX_i=\bx, U_i=u]}.
$$
We simulated the actual exposure status for each individual $A_i$ using a logistic regression model, 
\begin{equation}
\label{eqn:simAmodel}
\Pr(A_i=1|\bX_i=\bx) = \frac{\exp(\phi + \boldsymbol{\psi}^T \bx)}{1 + \exp(\phi + \boldsymbol{\psi}^T \bx)}.
\end{equation}
The observed outcome $Y$ was obtained for each individual by setting $Y_i = A_i Y_i(1) + (1-A_i) Y_i(0)$s. For each individual $i$, the observed data were ($A_i, \boldsymbol{X}_i, Y_i$). 

\vspace{-0.3cm}	
\subsection{Additional details about the simulation studies}
\label{AppSec:sim_studies}

As indicated in the main text, in each of the studies and scenarios described below, we first estimated $\bpi(a,\bx;\beeta_1)$ using a multinomial regression model and $e(\bx)$ using a logistic regression model. Then, we computed the estimators for the conditional statistical estimand, the total effect, and the three $\sface$ estimators.

Three studies were investigated. In Study I, we fixed the model parameters and varied the sample size between 5,000 and 50,000. In Study II, we varied $\delta_2$, the effect of the unmeasured covariate $U$ on $Y^{(2)}$, while keeping all the other parameters fixed. In Study III, we investigated the estimators' performance when the models fitted were misspecified. To achieve this, a transformation of $X_2$ was used to create the data, but the analysis assumed $X_2$ is (logit) linearly connected to the exposure or/and the outcome. Misspecification of the $A \sim X$ relationship was created by taking $\log|X_2|$ when generating the data instead of $X_2$. Misspecification of the $Y \sim A+X$ model was also created by replacing $X_2$ with $\log|X_2|$ when generating the data. Table \ref{Tab:params_for_simulation} gives the model parameters and sample sizes used in the three studies. It also gives the obtained true $\sface$ values for each set of parameters yields. In all studies, the prevalence of subtype 1 ranged between 5\%-6\% and the prevalence of subtype 2 ranged between 2\%-4\%. The prevalence of the exposure was 49\%.

\begin{sidewaystable}
\centering
\scriptsize
\renewcommand{\arraystretch}{2}
\resizebox{\textwidth}{!}{\begin{tabular}{cccccccccccccccc}
Study & n & $\alpha_1$ & $\alpha_2$ & $\beta_1$ & $\beta_2$ & $\bgamma_1$ & $\bgamma_2$ & $\delta_1$ & $\delta_2$ & $\phi$ & $\bpsi$ & $\sface^1_D$$^\dagger$ & $\sface^2_D$$^\dagger$ & $\sface^1_{RR}$ & $\sface^2_{RR}$\\ 
  \hline\\ 
 I & 5,000 & $\log(0.05)$  & $\log(0.005)$  & $\log(2)$   & $\log(2)$   & ($\log(0.25)$, $\log(2)$) & ($\log(2)$, $\log(2)$) & $\log(2)$  & $\log(2)$  & $\log(0.7)$  & ($\log(2)$, $\log(2)$) &  3470.5 &  969.1 & 1.75 & 1.61 \\ 
 & 10,000 & $\log(0.05)$  & $\log(0.005)$  & $\log(2)$   & $\log(2)$   & ($\log(0.25)$, $\log(2)$) & ($\log(2)$, $\log(2)$) & $\log(2)$  & $\log(2)$  & $\log(0.7)$  & ($\log(2)$, $\log(2)$) &  3470.5 &  969.1 & 1.75 & 1.61\\ 
 & 25,000 & $\log(0.05)$  & $\log(0.005)$  & $\log(2)$   & $\log(2)$   & ($\log( 0.25)$, $\log(2)$)& ($\log(2)$, $\log(2)$) & $\log(2)$  & $\log(2)$  & $\log(0.7)$  & ($\log(2)$, $\log(2)$) &  3470.5 &  969.1 & 1.75 & 1.61\\
 & 50,000 & $\log(0.05)$  & $\log(0.005)$  & $\log(2)$   & $\log(2)$   & ($\log(0.25)$, $\log(2)$) & ($\log(2)$, $\log(2)$) & $\log(2)$  & $\log(2)$  & $\log(0.7)$  & ($\log(2)$, $\log(2)$) &  3470.5 &  969.1 & 1.75 & 1.61\\ 
  II & 10,000 & $\log(0.05)$  & $\log(0.005)$  & $\log(2)$   & $\log(2)$   & ($\log(0.25)$, $\log(2)$) & ($\log(2)$, $\log(2)$) & $\log(2)$  & $\log(2)$  & $\log(0.7)$  & ($\log(2)$, $\log(2)$) &  3470.5 &  969.1 & 1.75 & 1.61\\ 
 & 10,000 & $\log(0.05)$  & $\log(0.005)$  & $\log(2)$   & $\log(2)$   & ($\log(0.25)$, $\log(2)$) & ($\log(2)$, $\log(3)$) & $\log(2)$  & $\log(2)$  & $\log(0.7)$  & ($\log(2)$, $\log(2)$) & 3417.6 & 1228.2 & 1.78 & 1.76\\  
 & 10,000 & $\log(0.05)$  & $\log(0.005)$  & $\log(2)$   & $\log(2)$   & ($\log(0.25)$, $\log(2)$)& ($\log(2)$, $\log(4)$) & $\log(2)$  & $\log(2)$  & $\log(0.7)$  & ($\log(2)$, $\log(2)$) & 3364.5 & 1498.5 & 1.77 & 1.71\\ 
 & 10,000 & $\log(0.05)$  & $\log(0.005)$  & $\log(2)$   & $\log(2)$   & ($\log(0.25)$, $\log(2)$) & ($\log(2)$, $\log(5)$) & $\log(2)$  & $\log(2)$  & $\log(0.7)$  & ($\log(2)$, $\log(2)$) & 3320.2 & 1727.2 & 1.76 & 1.65\\  
 & 10,000 & $\log(0.05)$  & $\log(0.005)$  & $\log(2)$   & $\log(2)$   & ($\log(0.25)$, $\log(2)$) & ($\log(2)$, $\log(6)$) & $\log(2)$  & $\log(2)$  & $\log(0.7)$  & ($\log(2)$, $\log(2)$) & 3261.1 & 1947.6 & 1.76 & 1.61\\ 
 III & 10,000 & $\log(0.05)$  & $\log(0.005)$  & $\log(2)$   & $\log(2)$   & ($\log(0.25)$, $\log(2)$) & ($\log(2)$, $\log(2)$) & $\log(2)$  & $\log(2)$  & $\log(0.7)$  & ($\log(2)$, $\log(2)$) & 2406.1 to 3470.5 & 656.6 to 969.1 & 1.63 to 1.78 & 1.67 to 1.81 \\ 
\end{tabular}}
\caption{Parameter values used in the simulations. $^\dagger$: effect per 100,000 people. Study III consists of different misspecified models that yield different effects, so the true effects are given in ranges.}
\label{Tab:params_for_simulation}
\end{sidewaystable}

\vspace{-0.5cm}	
\subsection{Simulation Results}
\label{AppSec:sim_res}

Tables \ref{Tab:SimsStudy1sfaces} and \ref{Tab:SimsStudy1theta} present the result for Study I. As can be seen from the Tables, the $\sface$ estimators had a smaller relative bias than the other estimators, even with a small sample size. The SD of all estimators in each sample size was similar, and the bootstrap SE estimator was approximately unbiased. The Wald-type CIs for the $\sface$ estimators had a satisfactory empirical coverage level, close to the desired 95\%.

\begin{table}[H]
\centering
\resizebox{\textwidth}{!}{\begin{tabular}{llll|lrrrr}
\multirow{2}{*}{Subtype} & \multirow{2}{*}{Sample Size} & 
\multirow{2}{*}{Effect} &
\multirow{2}{*}{Method}& \multicolumn{5}{c}{Diff} \\
&&& &Bias & \%Bias & CP95 & emp.SD & est.SE \\ 
  \hline
subtype 1 & 5,000 & $\sface$ & Stand & -18.3 & -0.5 & 94.6 & 694.6 & 703.0 \\ 
&&& IPTW & -26.2 & -0.8 & 94.8 & 711.1 & 718.1 \\ 
&&& DR & -23.0 & -0.7 & 94.9 & 704.5 & 714.2 \\ 
&& $TE$ & Stand  & -86.8 & -2.5 & 94.5 & 681.1 & 689.2 \\ 
&& $Cond$& Stand  & 98.3 & 2.8 & 94.7 & 706.7 & 713.7 \\ 
&10,000 & $\sface$ & Stand & 1.9 & 0.1 & 94.9 & 492.0 & 496.2 \\ 
  &&& IPTW & -3.0 & -0.1 & 94.9 & 502.5 & 507.4 \\ 
 &&& DR & -1.9 & -0.1 & 94.9 & 499.3 & 504.6 \\ 
  && $TE$& Stand  & -67.1 & -1.9 & 95.1 & 482.4 & 486.5 \\ 
   && $Cond$ & Stand & 119.3 & 3.4 & 94.7 & 499.3 & 503.7 \\ 
 & 25,000 & $\sface$ & Stand & 10.1 & 0.3 & 96.2 & 309.2 & 315.3 \\ 
 &&& IPTW & 4.1 & 0.1 & 95.5 & 320.9 & 322.5 \\ 
 &&& DR & 4.9 & 0.1 & 95.3 & 318.5 & 320.8 \\ 
   && $TE$& Stand  & -59.4 & -1.7 & 95.7 & 303.2 & 309.1 \\ 
   && $Cond$ & Stand & 129.3 & 3.7 & 93.9 & 314.0 & 320.1 \\ 
 & 50,000 & $\sface$ & Stand & 10.8 & -0.3 & 95.1 & 221.2 & 221.3 \\ 
 &&& IPTW & -14.1 & -0.4 & 95.2 & 225.0 & 226.1 \\ 
 &&& DR & -13.7 & -0.3 & 95.0 & 225.0 & 225.0 \\ 
   && $TE$& Stand  & -79.7 & -2.3 & 93.8 & 217.1 & 217.1 \\ 
   && $Cond$ & Stand & 107.0 & 3.1 & 91.7 & 224.0 & 224.6 \\
  subtype 2 &  5,000 & $\sface$ & Stand & 6.3 & 0.6 & 94.4 & 393.4 & 393.0 \\ 
    &&& IPTW & 6.4 & 0.7 & 94.5 & 404.9 & 402.2 \\ 
    &&& DR & 5.5 & 0.6 & 94.9 & 402.0 & 401.7 \\ 
    && $TE$& Stand  & -69.0 & -7.1 & 93.9 & 362.9 & 362.7 \\ 
    && $Cond$& Stand  & 152.3 & 15.7 & 93.0 & 422.8 & 422.3 \\ 
  &10,000 & $\sface$ & Stand & 13.2 & 1.4 & 94.8 & 277.6 & 277.5 \\ 
    &&& IPTW & 12.7 & 1.3 & 94.6 & 286.0 & 285.2 \\ 
    &&& DR & 12.2 & 1.3 & 94.7 & 285.3 & 284.6 \\ 
   && $TE$& Stand  & -62.7 & -6.5 & 95.0 & 256.3 & 256.2 \\ 
    && $Cond$ & Stand & 160.2 & 16.5 & 91.2 & 298.9 & 298.1 \\ 
  &25,000 & $\sface$ & Stand & 23.1 & 2.4 & 93.5 & 182.7 & 175.9 \\ 
    &&& IPTW & 20.5 & 2.1 & 94.6 & 188.4 & 181.4 \\ 
    &&& DR & 20.5 & 2.1 & 94.4 & 188.8 & 180.9 \\ 
    && $TE$& Stand  & -53.5 & -5.5 & 93.5 & 168.7 & 162.3 \\ 
    && $Cond$& Stand  & 170.2 & 17.6 & 83.1 & 196.5 & 188.8 \\ 
  &50,000 & $\sface$ & Stand & 19.7 & 2.0 & 93.5 & 130.0 & 124.9 \\ 
    &&& IPTW & 17.4 & 1.8 & 93.5 & 135.4 & 128.4 \\ 
    &&& DR & 17.1 & 1.76 & 93.7 & 135.0 & 127.1 \\ 
    && $TE$& Stand  & -56.6 & -5.8 & 90.3 & 120.0 & 114.4 \\ 
    && $Cond$& Stand & 167.0 & 17.3 & 83.1 & 140.1 & 133.9 \\ 
\end{tabular} 
\quad \quad 
\begin{tabular}{lrrrr}
   \multicolumn{5}{c}{RR} \\
 Bias & \%Bias & CP95 & emp.SD & est.SE \\ 
  \hline
0.01 & 0.60 & 94.50 & 0.22 & 0.23 \\ 
0.01 & 0.56 & 94.40 & 0.23 & 0.24 \\ 
0.01 & 0.59 & 94.70 & 0.23 & 0.23 \\ 
0.01 & 0.60 & 94.50 & 0.22 & 0.23 \\ 
0.03 & 1.52 & 94.80 & 0.23 & 0.23 \\ 
 0.01 & 0.42 & 95.20 & 0.16 & 0.16 \\ 
0.01 & 0.35 & 95.70 & 0.16 & 0.16 \\ 
0.01 & 0.36 & 95.40 & 0.16 & 0.16 \\ 
0.01 & 0.42 & 95.20 & 0.16 & 0.16 \\ 
0.02 & 1.36 & 95.20 & 0.16 & 0.16 \\ 
0.01 & 0.29 & 95.20 & 0.10 & 0.10 \\   
0.00 & 0.13 & 94.70 & 0.10 & 0.10 \\  
0.00 & 0.15 & 94.80 & 0.10 & 0.10 \\ 
0.01 & 0.29 & 95.20 & 0.10 & 0.10 \\ 
0.02 & 1.25 & 95.70 & 0.10 & 0.10 \\
0.00 & -0.19 & 94.80 & 0.07 & 0.07 \\   
0.01 & -0.29 & 95.60 & 0.07 & 0.07 \\  
0.01 & -0.29 & 95.20 & 0.07 & 0.07 \\ 
0.00 & -0.19 & 94.80 & 0.07 & 0.07 \\ 
0.01 & 0.75 & 94.50 & 0.07 & 0.07 \\
0.11 & 6.26 & 95.70 & 0.54 & 5.55 \\ 
0.13 & 7.15 & 95.70 & 0.59 &  5.56 \\ 
0.13 & 7.00 & 96.50 & 0.59 &  5.55 \\ 
0.11 & 6.26 & 95.70 & 0.54 & 5.55 \\ 
0.19 & 10.61 & 96.50 & 0.57 & 5.55 \\ 
0.07 & 4.02 & 96.20 & 0.37 & 0.39 \\ 
0.08 & 4.47 & 96.10 & 0.39 & 0.41 \\ 
0.08 & 4.40 & 96.10 & 0.39 & 0.41 \\ 
0.07 & 4.02 & 96.20 & 0.37 & 0.39 \\ 
0.15 & 8.33 & 97.80 & 0.38 & 0.40 \\ 
0.05 & 2.83 & 95.20 & 0.23 & 0.23 \\ 
0.05 & 2.74 & 95.30 & 0.25 & 0.24 \\ 
0.05 & 2.74 & 95.00 & 0.25 & 0.24 \\ 
0.05 & 2.83 & 95.20 & 0.23 & 0.23 \\ 
0.13 & 7.01 & 95.10 & 0.24 & 0.24 \\ 
0.04 & 2.15 & 94.50 & 0.17 & 0.16 \\ 
0.04 & 2.07 & 93.80 & 0.18 & 0.17 \\ 
0.04 & 2.03 & 94.60 & 0.17 & 0.17 \\ 
0.04 & 2.15 & 94.50 & 0.17 & 0.16 \\ 
0.11 & 6.39 & 91.50 & 0.17 & 0.17 \\ 
\end{tabular}}
\caption{ \small Study I simulation results for $\sface^{(1)}$ and $\sface^{(2)}$. Three methods were considered to estimate the $\sface_{D}$; IPTW, standardization and DR. The $TE$ and $Cond$ were estimated using standardization. Diff: defined on the difference scale, RR: defined on the RR scale. \%Bias: relative bias, CP95: empirical coverage proportion of 95\% confidence intervals, emp.SD: empirical standard deviation of the simulation estimates, SE: mean estimated standard errors.}
\label{Tab:SimsStudy1sfaces}
\end{table}

\begin{table}[H]
\centering
\resizebox{\textwidth}{!}{\begin{tabular}{lll|lrrrr}
 \multirow{2}{*}{Sample Size} & 
\multirow{2}{*}{Effect} &
\multirow{2}{*}{Method}& \multicolumn{5}{c}{Diff} \\
&&& Bias & \%Bias & CP95 & emp.SD & est.SE \\ 
  \hline
5,000 & $\sface$ & Stand & -24.6 & -1.0 & 95.2 & 791.1 & 810.5 \\ 
&& IPTW & -32.6 & -1.3 & 94.5 & 809.1 & 828.8 \\ 
&& DR & -28.5 & -1.1 & 94.7 & 802.9 & 825.5 \\ 
& $TE$  & Stand & -17.7 & -0.7 & 95.1 & 771.5 & 790.2 \\ 
& $Cond$  & Stand & -54.0 & -2.2 & 95.3 & 790.5 & 808.8 \\ 
10,000 & $\sface$ & Stand &  -11.3 & -0.5 & 95.4 & 570.2 & 573.5 \\ 
&& IPTW & -15.7 & -0.6 & 95.3 & 583.8 & 587.3 \\ 
&& DR & -14.1 & -0.6 & 94.9 & 581.7 & 585.0 \\ 
& $TE$ & Stand & -4.4 & -0.2 & 95.3 & 555.7 & 559.2 \\ 
& $Cond$  & Stand & -40.9 & -1.6 & 95.1 & 569.6 & 572.4 \\ 
25,000 & $\sface$ & Stand &  -13.0 & -0.5 & 94.4 & 363.9 & 361.8 \\ 
&& IPTW & -16.4 & -0.7 & 94.1 & 378.2 & 370.9 \\ 
&& DR & -15.6 & -0.6 & 94.6 & 376.0 & 369.4 \\ 
& $TE$ & Stand & -5.9 & -0.2 & 94.2 & 354.4 & 352.8 \\ 
& $Cond$ & Stand & -40.9 & -1.6 & 94.2 & 362.9 & 361.0 \\ 
50,000 & $\sface$ & Stand &  -30.4 & -1.2 & 94.1 & 259.1 & 255.8 \\ 
&& IPTW & -31.5 & -1.3 & 94.7 & 264.4 & 261.1 \\ 
&& DR & -30.8 & -1.2 & 94.6 & 263.0 & 260.4 \\ 
& $TE$ & Stand & -23.1 & -0.9 & 94.5 & 252.1 & 248.1 \\ 
& $Cond$ & Stand & -60.1 & -2.4 & 94.0 & 258.0 & 254.1 \\ 
\end{tabular}
\quad \quad 
\begin{tabular}{rrrrr}
\multicolumn{5}{c}{RR} \\
 Bias & \%Bias & CP95 & emp.SD & est.SE \\ 
  \hline
 -0.10 & 374.86 & 97.10 & 0.58 & 0.68 \\ 
 -0.12 & 436.64 & 97.10 & 0.63 & 0.74 \\ 
 -0.12 & 424.36 & 97.30 & 0.63 & 0.74 \\ 
 -0.10 & 374.86 & 97.10 & 0.58 & 0.68 \\ 
 -0.17 & 602.76 & 97.40 & 0.59 & 0.70 \\ 
 -0.07 & 238.18 & 96.60 & 0.41 & 0.42 \\ 
-0.07 & 272.83 & 96.50 & 0.43 & 0.45 \\ 
-0.07 & 267.64 & 96.60 & 0.43 & 0.45 \\ 
-0.07 & 238.18 & 96.60 & 0.41 & 0.42 \\ 
-0.13 & 462.35 & 97.30 & 0.41 & 0.43 \\ 
-0.05 & 168.16 & 95.40 & 0.26 & 0.25 \\ 
-0.05 & 172.42 & 95.20 & 0.27 & 0.26 \\ 
-0.05 & 171.47 & 95.30 & 0.27 & 0.26 \\ 
-0.05 & 168.16 & 95.40 & 0.26 & 0.25 \\ 
-0.10 & 382.06 & 95.20 & 0.26 & 0.25 \\
-0.04 & 154.71 & 94.70 & 0.18 & 0.17 \\ 
-0.04 & 155.81 & 94.90 & 0.19 & 0.18 \\ 
-0.04 & 153.37 & 94.80 & 0.19 & 0.18 \\ 
-0.04 & 154.71 & 94.70 & 0.18 & 0.17 \\ 
-0.10 & 373.38 & 93.30 & 0.18 & 0.18 \\
\end{tabular}}
\caption{ \small Study I simulation results for $\theta$. Three methods were considered to estimate the $\sface$; IPTW, standardization and DR. The $TE$ and $Cond$ were estimated using standardization. Diff: defined on the difference scale, RR: defined on the RR scale. \%Bias: relative bias, CP95: empirical coverage proportion of 95\% confidence intervals, emp.SD: empirical standard deviation of the simulation estimates, SE: mean estimated standard errors.}
\label{Tab:SimsStudy1theta}
\end{table}

\begin{figure}[H]
\includegraphics[scale=0.5]{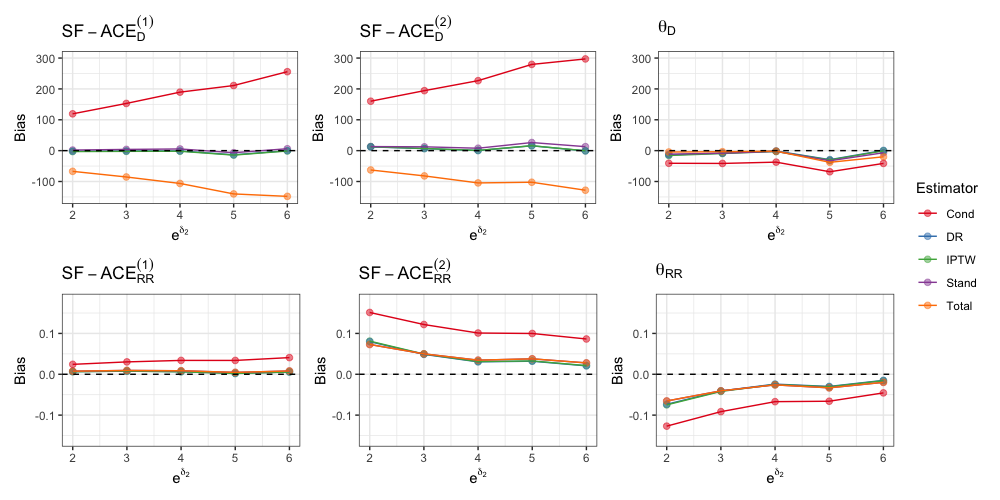}
\caption{Results for Study II. Absolute bias for $\sface^{(1)}$, $\sface^{(2)}$ and $\theta$, as a function of  the effect of $U$ on $Y^{(2)}$ $(e^{\delta_2})$. For the difference scale, results are presented per 100,000 people. $D$: defined on the difference scale, $RR$: defined on the RR scale.}
\label{Fig:Upar_bias}
\end{figure}

\begin{figure}[H]
\includegraphics[scale=0.5]{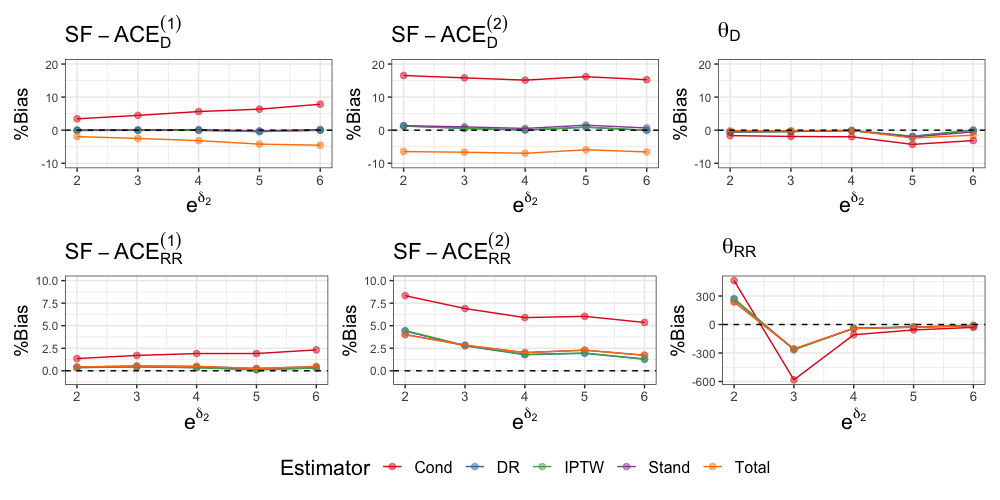}
\caption{Results for Study II. Relative bias for $\sface^{(1)}$, $\sface^{(2)}$ and $\theta$, as a function of  the effect of $U$ on $Y^{(2)}$ $(e^{\delta_2})$. $D$: defined on the difference scale, $RR$: defined on the RR scale.}
\label{Fig:Upar_relative_bias}
\end{figure}

Turning to Study II, Figures \ref{Fig:Upar_bias} and \ref{Fig:Upar_relative_bias}   presents the bias of the different estimators, relative to the true $\sface$s and $\theta$, and as a function of the strength of the effect of $U$ on $Y^{(2)}$ $(\exp(\delta_2))$. The absolute difference between non-$\sface$ estimators and the $\sface$s on the difference scale increased (in absolute value) as  increased and remained the similar on the RR scale (Figure \ref{Fig:Upar_bias}). As expected, because the data were simulated under \Smono for both subtypes the $TE$ and $\sface$ coincide on the RR scale. The relative bias (Figure \ref{Fig:Upar_relative_bias}) remained generally the same. As all three methods (standardization, IPTW, and DR estimation) were based on correctly-specified models, their finite sample bias was minimal.

Finally, Table \ref{Tab:Study3miss} presents results from Study III under four settings: (a) Both $\bpi(a,\bx;\beeta_1)$ and $e(\bx;\beeta_2)$ are correctly specified, (b) The $\bpi(a,\bx;\beeta_1)$ model is correctly-specified and model $e(\bx;\beeta_2)$ is misspecified, (c) The $\bpi(a,\bx;\beeta_1)$ model is misspecified and model $e(\bx;\beeta_2)$ is correctly-specified, and (d) both models are misspecified. As expected the standardization-based estimator had minimal bias when $\bpi(a,\bx;\beeta_1)$ was correctly specified, and was biased when the model was misspecified. The IPTW-based model had minimal bias whenever model $e(\bx;\beeta_2)$ was correctly specified and produced considerable bias when the model was  misspecified. As expected the DR estimator was only biased when both models were misspecified and had minimal bias under scenarios (a), (b), and (c) described above.

Table \ref{Tab:Study3miss} also shows that the SE was well estimated by the bootstrap and that the Wald-type 95\% confidence intervals had satisfactory coverage rate whenever the respective model was correctly-specified.
	
\begin{table}[H]
\centering
\begin{tabular}{lrrrrrr}
Method & Bias & \%Bias & CP95 & emp.SD & est.SE \\ 
  \hline
  \bf{Both models correct} &&&& \\[1ex]
~~Stand  & 1.88 & 0.1& 94.9 & 492.0 & 496.2 \\ 
~~IPTW  & -2.99 & -0.1 & 94.9 & 502.5 & 507.3 \\ 
~~DR & -1.91 & -0.1 & 94.9 & 499.3 & 504.6 \\ 
\bf{Model $A \sim X$ is misspecified} &&&& \\[1ex]
~~Stand  & -35.05 & -1.0 & 95.0 & 508.2 & 509.6 \\ 
~~IPTW  & 557.64 & 16.1 & 81.1 & 510.3 & 516.1 \\
~~DR  & -37.65 & -1.1 & 94.8 & 508.3 & 510.0 \\ 
\bf{Model $Y \sim A+X$ is misspecified} &&&& \\[1ex]
~~Stand  & -182.67 & -7.6 & 91.6 & 419.5 & 406.9 \\ 
~~IPTW  & -4.19 & -0.2 & 93.6 & 494.8 & 481.0 \\ 
~~DR  & -4.56 & -0.2 & 93.9 & 489.2 & 476.0 \\ 
\bf{Both models misspecified} &&&& \\
~~Stand  & 2344.12 & 97.4 & 0.2 & 451.9 & 473.3 \\ 
~~IPTW  & 2344.82 & 97.5 & 0.2 & 450.8 & 473.7 \\ 
~~DR  & 2338.78 & 97.2 & 0.2 & 451.3 & 473.9 \\ 
\end{tabular}
\caption{Results for Study III. Performance of the $\sface^{(1)}$ estimators when different models are correctly-specified and/or misspecified. Results are presented per 100,000 people. Three methods were considered to estimate the $\sface_{D}$; IPTW, standardization and DR. The $TE_{D}$ was estimated using standardization. \%Bias: relative bias, CP95: Empirical coverage proportion of 95\% confidence intervals, emp.SD: empirical standard deviation of the simulation estimates, SE: mean estimated standard errors.}
\label{Tab:Study3miss}
\end{table}
 
\begin{table}[H]
\centering
\begin{tabular}{lrrrrrr}
Method & Bias & \%Bias & CP95 & emp.SD & est.SE \\ 
  \hline
  \bf{Both models correct} &&&& \\[1ex]
~~Stand & 13.16 & 1.36 & 94.80 & 277.60 & 277.52 \\ 
~~IPTW & 12.68 & 1.31 & 94.60 & 285.99 & 285.18 \\ 
~~DR & 12.18 & 1.26 & 94.70 & 285.28 & 284.61 \\ 
\bf{Model $A \sim X$ is misspecified} &&&& \\[1ex]
~~Stand & 24.18 & 2.49 & 94.70 & 276.21 & 274.79 \\ 
 ~~IPTW  & 199.86 & 20.62 & 89.40 & 279.62 & 278.17 \\ 
~~DR & 23.02 & 2.38 & 95.00 & 275.89 & 275.13 \\ 
 \bf{Model $Y \sim A+X$ is misspecified} &&&& \\[1ex]
~~Stand  & -89.72 & -13.6 & 92.6 & 217.5 & 214.5\\ 
~~IPTW  & 9.47 & 1.4 & 94.8 & 250.1 & 250.2 \\ 
~~DR  & 9.26 & 1.4 & 94.6 & 249.7 & 249.5 \\ 
\bf{Both models misspecified} &&&& \\
~~Stand  & 659.26 & 100.4 & 21.2 & 247.2 & 243.5 \\ 
~~IPTW  & 663.16 & 100.9 & 21.2 & 246.6 & 243.6 \\ 
~~DR  & 656.57 & 99.9 & 21.6 & 246.9 & 243.9 \\ 
   \hline
\end{tabular}
\caption{Results for Study III. Performance of the $\sface^{(2)}$ estimators when different models are correctly-specified and/or misspecified. Results are presented per 100,000 people. Three methods were considered to estimate the $\sface_{D}$; IPTW, standardization and DR. The $TE_{D}$ was estimated using standardization. \%Bias: relative bias, CP95: Empirical coverage proportion of 95\% confidence intervals, emp.SD: empirical standard deviation of the simulation estimates, SE: mean estimated standard errors.}
\label{Tab:Study3missSubtype2}
\end{table}
\section{Details on the data analysis and additional results}
\label{AppSec:AppData}
Section \ref{AppSec:data_info} presents details on the data  as well as additional analyses and results. Section \ref{AppSec:data_info} includes descriptive information about the data used. Section \ref{AppSec:data_models_info} gives additional information about the weights used for handling missing subtypes and about the models fitted for estimating effects. Finally, Section \ref{AppSec:data_res_info} presents additional  data analysis results.

\subsection{Additional information about the data}
\label{AppSec:data_info}
Table \ref{Tab:descriptive} gives basic descriptive information about the different covariates $\bX$ used in our analyses. 

\begin{table}[!htbp]
\centering
\small
\captionsetup{font=footnotesize}
\resizebox{\textwidth}{!}{\begin{tabular}{llllll}
  & Total & CRC free & MSI-high & Non-MSI-high & Missing subtype\\
 & n = 109,272 & n = 108,853 & n = 61 & n = 358 & n=542\\
 \hline
Smoking &  &  &  & &\\
~~No & 50840 (46.5\%) & 50667 (46.5\%) & 24 (39.3\%) & 149 (41.6\%) & 229 (42.3\%)\\
~~Yes & 58432 (53.5\%) & 58186 (53.5\%) & 37 (60.7\%) & 209 (58.4\%) & 313 (57.7\%)\\[1ex]
Gender &  &  &  & &\\
~~Men (HPFS) & 28801 (26.4\%) & 28675 (26.3\%) & 15 (24.6\%) & 111 (31\%)  & 169 (31.2\%)\\
~~Women (NHS) & 80471 (73.6\%) & 80178 (73.7\%) & 46 (75.4\%) & 247 (69\%)  & 373 (68.8\%)\\[1ex]
Body mass index &  &  &  & &\\
~~high & 14506 (13.3\%) & 14440 (13.3\%) & 9 (14.8\%) & 57 (15.9\%) & 97 (17.9\%)\\
~~low & 56789 (52\%) & 56607 (52\%) & 26 (42.6\%) & 156 (43.6\%) & 253 (46.7\%)\\
~~normal & 37977 (34.8\%) & 37806 (34.7\%) & 26 (42.6\%) & 145 (40.5\%) & 192 (35.4\%)\\[1ex]
Using aspirin regularly &  &  &  & &\\
~~No & 71180 (65.1\%) & 70898 (65.1\%) & 43 (70.5\%) & 239 (66.8\%) & 371 (68.5\%)\\
~~Yes & 38092 (34.9\%) & 37955 (34.9\%) & 18 (29.5\%) & 119 (33.2\%) & 171 (31.5\%)\\[1ex]
Family history of colorectal cancer &  &  &  & &\\
~~No & 95990 (87.8\%) & 95641 (87.9\%) & 48 (78.7\%) & 301 (84.1\%) & 469 (86.5\%)\\
~~Yes & 13282 (12.2\%) & 13212 (12.1\%) & 13 (21.3\%) & 57 (15.9\%) & 73 (13.5\%)\\[1ex]
History of endoscopy &  &  &  & &\\
~~No & 67821 (62.1\%) & 67511 (62\%) & 41 (67.2\%) & 269 (75.1\%)& 410 (75.6\%)\\
~~Yes & 41451 (37.9\%) & 41342 (38\%) & 20 (32.8\%) & 89 (24.9\%) & 132 (24.4\%)\\[1ex]
Alcohol intake, g/d 
 & 7.4 (11.0) & 7.4 (11.0) & 6.7 (10.6) & 8.3 (12.5) & 9.0 (13.3)\\[2ex]
Total calorie intake, ($kcal/day$) 
 & 1762.4 (496.5) & 1762.4 (496.5) & 1695.2 (460.5) & 1767.9 (499.3) & 1748.9 (542.5)\\[2ex]
Physical activity, METS-h/wk &  20.0 (18.7) & 20.0 (18.7) & 17.3 (14.2) & 18.7 (20.2) & 19.0 (16.2)
 \end{tabular}}
\caption{Descriptive information about baseline covariates and the exposure in the data. Values are means (SD) for continuous variables and counts (percentages) for categorical variables. Total includes participants left in the sample after removing participants with missing subtype.}
\label{Tab:descriptive}
\end{table}

\subsection{Additional information about the models}
\label{AppSec:data_models_info}

\begin{figure}[H]
\centering
\includegraphics[scale=0.45]{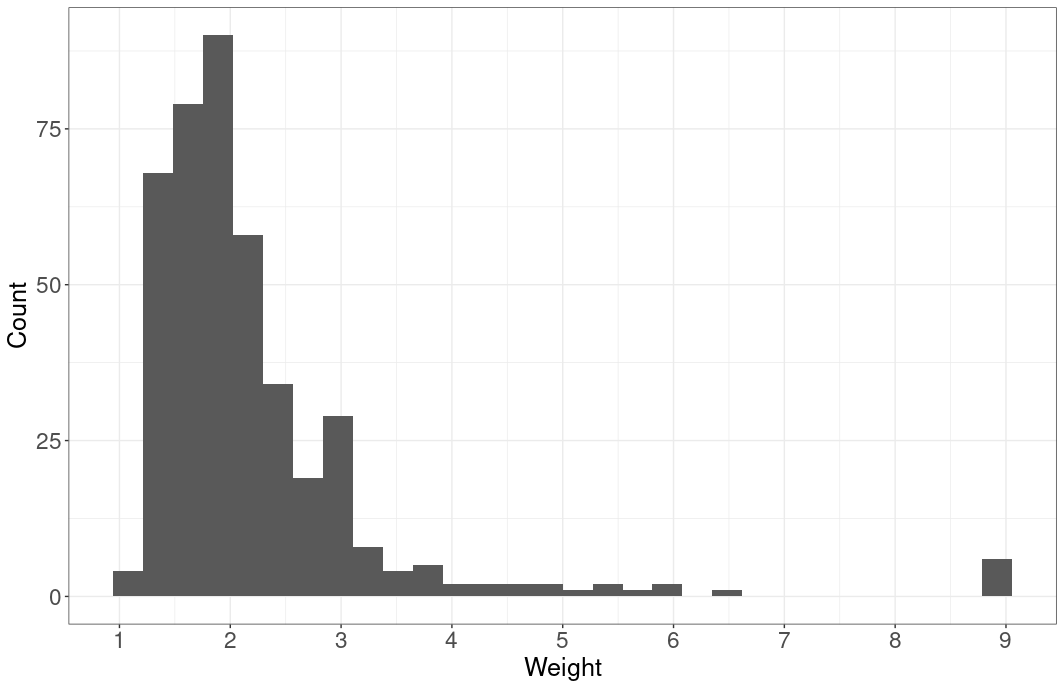}
\captionsetup{font=footnotesize}
\caption{Distribution of the inverse probability weights for missing subtype.}
\label{Fig:hist_missing_subtype}
\end{figure}

To take into account the individuals diagnosed with CRC without having their MSI subtype identified, we used inverse probability weighting \citep{liu2018utility}. We fitted a logistic regression model among the individuals diagnosed with CRC, the outcome being missing or available MSI subtype. The covriates included were tumor location (proximal colon, distal colon, rectum, missing) disease stage (stage I, stage II, stage III, stage IV, missing), tumor differentiation (well, moderate, poor, unspecified, missing), age of diagnosis, year of diagnosis (1976–1995, 1996–2000, 2001–2012) and family history of colorectal cancer (yes or no). The estimates and 95\% CI of the model coefficients are given in Table \ref{table:logistic_summary_missing_subtype}. After fitting the model, we removed the individuals who were diagnosed with CRC but their MSI subtype was missing, and used weights based on the inverse of the logistic regression predictions to adjust our sample. We truncated the weights at the 99\% percentile to reduce variance. The distribution of the final weights used are given in Figure \ref{Fig:hist_missing_subtype}.

\begin{table}[H] \centering 
\begin{tabular}{@{\extracolsep{5pt}}lll} 
Covariate & Estimate & 95\% CI \\ 
\hline \\
 Gender & 0.31 & (0.04, 0.58) \\ 
  Tumor location - missing & $-$2.19 &($-$3.61, $-$0.76) \\ 
  Tumor location - proximal colon & 0.02 &($-$0.24, 0.29) \\ 
  Tumor location - rectum & $-$0.18& ($-$0.51, 0.14) \\ 
  Disease stage - II & 0.01 &($-$0.32, 0.33) \\ 
  Disease stage - III & $-$0.05 &($-$0.39, 0.29) \\ 
  Disease stage - IV & $-$0.49 &($-$0.87, $-$0.11) \\ 
  Disease stage - missing & $-$1.50& ($-$1.96, $-$1.03) \\ 
  Tumor differentiation - missing & $-$0.50 &($-$2.35, 1.35) \\ 
  Tumor differentiation - poor & 0.06 &($-$0.26, 0.38) \\ 
  Tumor differentiation - unspecified & $-$0.59 &($-$0.96, $-$0.21) \\ 
  Tumor differentiation - well & $-$0.26 &($-$0.63, 0.11) \\ 
  Age of diagnosis & 0.03 &($-$0.002, 0.06) \\ 
  Year of diagnosis - 2001–2012 & $-$0.66 &($-$0.96, $-$0.37) \\ 
  Year of diagnosis - 1976–1995 & $-$0.65 &($-$0.98, $-$0.33) \\
  Family history of CRC & 0.17 &($-$0.15, 0.49) \\ 
\end{tabular} 
\caption{Estimates and 95\% Wald-type confidence intervals for the coefficients of the logistic regression model used for calculating weights for missing subtype. } 
\label{table:logistic_summary_missing_subtype}
\end{table}

When estimating the $\sface$ using IPTW and DR estimators, we fitted a logistic regression model, predicting the probability of smoking by age 60 based on the baseline covariats. The summary of this model is given in Table \ref{table:logistic_summary}. Similarly, when estimating the $\sface$ and $TE$ using standardization and DR estimators, we fitted a multinomial regression model, predicting the probability of having non-MSI-high CRC, MSI-high CRC or being CRC free by age 70, based on smoking status by age 60 and the baseline covariates. The summary of this model is given in Table \ref{table:multinomial_summary}.

\begin{table}[H]
\centering
\resizebox{\textwidth}{!}{\begin{tabular}{@{\extracolsep{5pt}}lcccc} 
 Covariate & \multicolumn{2}{c}{non-MSI-high} & \multicolumn{2}{c}{MSI-high} \\  
 \hline 
 & Estimated OR & 95\%CI & Estimated OR & 95\%CI \\[1ex] 
Smoking & 1.19 &(1.06, 1.33) & 1.39$^{**}$ &(1.38, 1.41) \\
Gender (Men) & 1.25$^{**}$ &(1.08, 1.42) & 0.87$^{**}$ &(0.83, 0.92) \\
Body mass index (low) & 0.68$^{**}$ &(0.54, 0.83) & 0.81$^{*}$ &(0.61, 1.01) \\ Body mass index (normal) & 1.04 &(0.90, 1.19) & 1.20$^{*}$ &(1.00, 1.39) \\ Using aspirin regularly & 0.93 &(0.78, 1.08) & 0.86$^{**}$ &(0.84, 0.87) \\ Alcohol intake & 1.00 &(1.00, 1.01) & 1.00 &(0.98, 1.02) \\  Total calorie intake & 1.00 &(1.00, 1.00) & 1.00$^{**}$ &(1.00, 1.00) \\ Family history of CRC & 1.52$^{**}$ & (1.33, 1.71) & 1.89$^{**}$ & (1.89, 1.90) \\ 
History of endoscopy & 0.44$^{**}$ &(0.27, 0.61) & 0.81$^{**}$ &(0.80, 0.81) \\ Physical activity & 1.00 &(1.00, 1.00) & 0.99 &(0.98, 1.01) \\ 
\end{tabular}} 
\caption{Estimates and 95\% Wald-type confidence intervals for the odds ratio obtained from the multinomial regression model used for standardization and DR estimation. $^{*}$p$<$0.05, $^{**}$p$<$0.01}
 \label{table:multinomial_summary}
\end{table}

\begin{table}[H] \centering 
\begin{tabular}{@{\extracolsep{5pt}}lcc} 
Covariate & Estimated OR & 95\%CI\\ 
\hline
 Gender (Men) & 0.67$^{**}$ & (0.64, 0.71) \\ 
  Body mass index (low) & 0.99& (0.95, 1.02) \\ 
  Body mass index (normal) & 1.04& (1.00, 1.08) \\ 
  Using aspirin regularly & 1.05$^{**}$& (1.03, 1.08) \\ 
  Alcohol intake & 1.05$^{**}$ & (1.05, 1.05) \\ 
  Total calorie intake & 1.00$^{**}$& (1.00, 1.00) \\ 
  Family history of CRC & 1.00 & (0.96, 1.04) \\ 
  History of endoscopy & 0.95$^{**}$ &(0.92, 0.98) \\ 
   Physical activity & 1.00$^{**}$ &(1.00, 1.00) \\ 
\end{tabular} 
 \caption{Estimates and 95\% Wald-type confidence intervals for the odds ratio obtained from the logistic regression model used for IPTW and DR estimation. OR: odds ratio $^{*}$p$<$0.05, $^{**}$p$<$0.01}
 \label{table:logistic_summary}
\end{table}

\subsection{Additional results}
\label{AppSec:data_res_info}

Turning to the analysis results. First, we estimated the effects under \Smono for both disease subtypes. The results for the $\sface_{D}$ are given in Table 2 of the main text. Here, we give the results for the $\sface_{RR}$ in Table \ref{Tab:data_res_Smono_RR}, and the results for $\theta$ in Table \ref{Tab:data_res_theta_Smono_RR}.

\begin{figure}[H]
\centering
\includegraphics[scale=0.45]{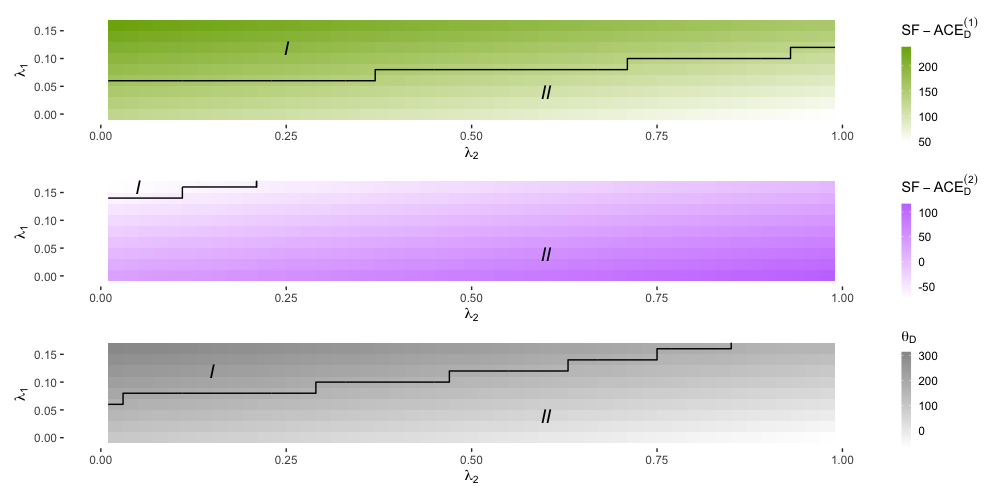}
\captionsetup{font=footnotesize}
\caption{Sensitivity analysis for estimating $\sface_D^{(1)}$, $\sface_D^{(2)}$ and $\theta_{D}$ under \Dmono for both MSI CRC subtypes. The figure presents standardization estimates as a function of $\lambda_1= \Pr[Y^{(2)}(1) = 1|Y^{(1)}(0) = 1]$ and $\lambda_2 = \Pr[Y^{(1)}(1) = 1|Y^{(2)}(0) = 1]$. The black lines divide the grid such that in sections marked with $I$ the effect is significant in 5\% level while in sections marked with $II$ it is not.}
\label{Fig:diff_sens2_stand}
\end{figure}

\begin{figure}[H]
\centering
\includegraphics[scale=0.45]{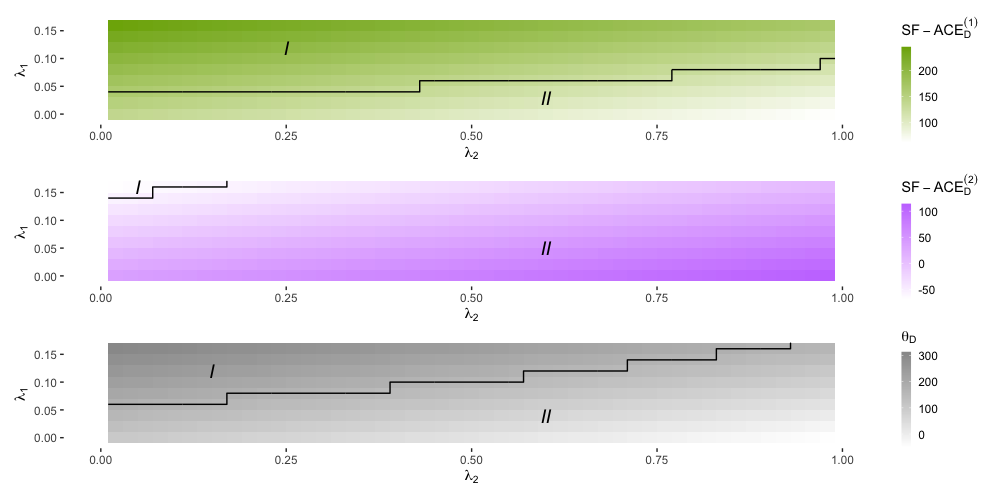}
\captionsetup{font=footnotesize}
\caption{Sensitivity analysis for estimating $\sface_D^{(1)}$, $\sface_D^{(2)}$ and $\theta_{D}$ under \Dmono for both MSI CRC subtypes. The figure presents IPTW estimates as a function of $\lambda_1= \Pr[Y^{(2)}(1) = 1|Y^{(1)}(0) = 1]$ and $\lambda_2 = \Pr[Y^{(1)}(1) = 1|Y^{(2)}(0) = 1]$. The black lines divide the grid such that in sections marked with $I$ the effect is significant in 5\% level while in sections marked with $II$ it is not.}
\label{Fig:diff_sens2_iptw}
\end{figure}

\begin{figure}[H]
\includegraphics[width=\textwidth,height=\textheight,keepaspectratio]{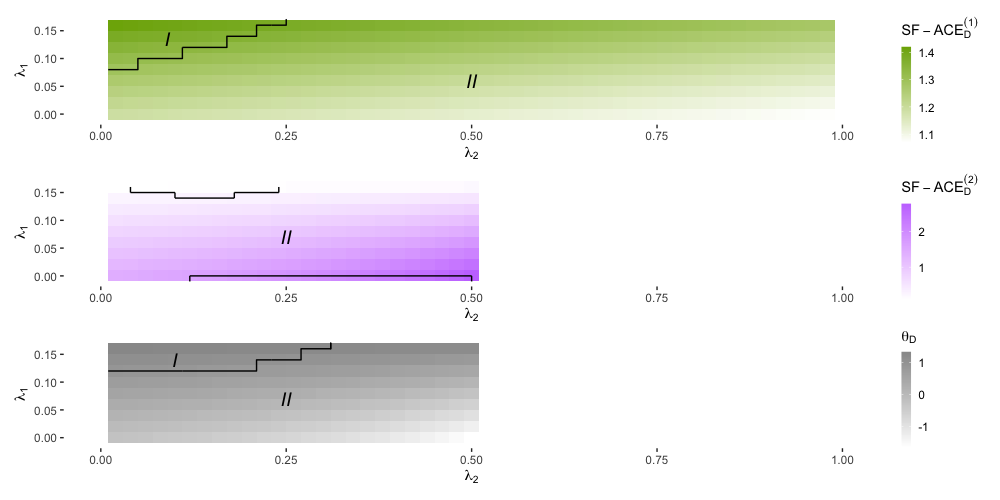}
\captionsetup{font=small}
\caption{Sensitivity analysis for estimating $\sface_{RR}^{(1)}$, $\sface_{RR}^{(2)}$ and $\theta_{RR}$ under \Dmono for both MSI CRC subtypes. The figure presents standardization estimates as a function of $\lambda_1= \Pr[Y^{(2)}(1) = 1|Y^{(1)}(0) = 1]$ and $\lambda_2 = \Pr[Y^{(1)}(1) = 1|Y^{(2)}(0) = 1]$. The black lines divide the grid such that in sections marked with $I$ the effect is significant in 5\% level while in sections marked with $II$ it is not.}
\label{Fig:RR_sens2_stand}
\end{figure}

\begin{figure}[H]
\includegraphics[width=\textwidth,height=\textheight,keepaspectratio]{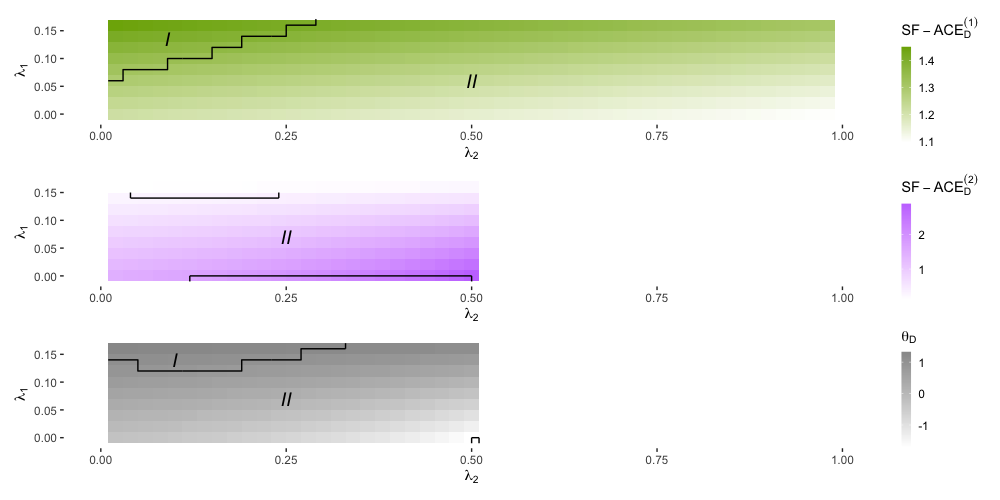}
\captionsetup{font=small}
\caption{Sensitivity analysis for estimating $\sface_{RR}^{(1)}$, $\sface_{RR}^{(2)}$ and $\theta_{RR}$ under \Dmono for both MSI CRC subtypes. The figure presents IPTW estimates as a function of $\lambda_1= \Pr[Y^{(2)}(1) = 1|Y^{(1)}(0) = 1]$ and $\lambda_2 = \Pr[Y^{(1)}(1) = 1|Y^{(2)}(0) = 1]$. The black lines divide the grid such that in sections marked with $I$ the effect is significant in 5\% level while in sections marked with $II$ it is not.}
\label{Fig:RR_sens2_iptw}
\end{figure}

\begin{figure}[H]
\includegraphics[width=\textwidth,height=\textheight,keepaspectratio]{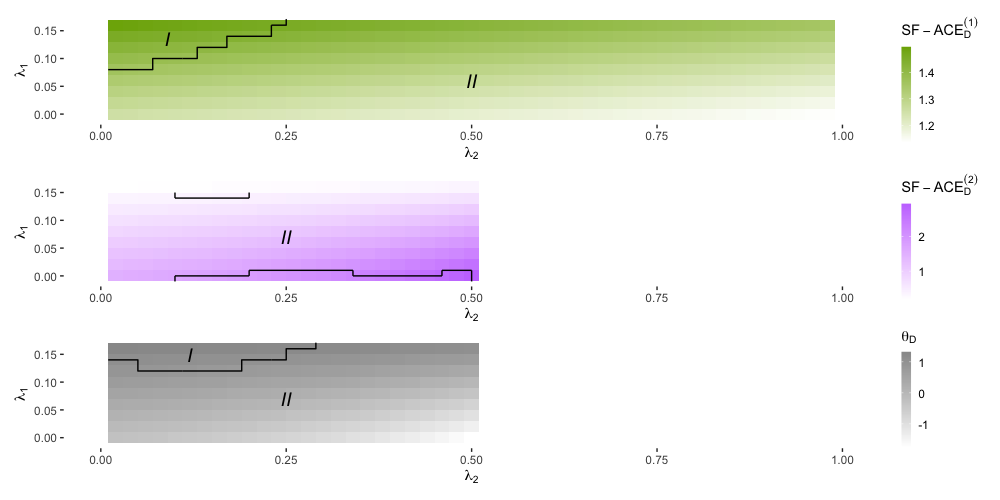}
\captionsetup{font=small}
\caption{Sensitivity analysis for estimating $\sface_{RR}^{(1)}$, $\sface_{RR}^{(2)}$ and $\theta_{RR}$ under \Dmono for both MSI CRC subtypes. The figure presents DR estimates as a function of $\lambda_1= \Pr[Y^{(2)}(1) = 1|Y^{(1)}(0) = 1]$ and $\lambda_2 = \Pr[Y^{(1)}(1) = 1|Y^{(2)}(0) = 1]$. The black lines divide the grid such that in sections marked with $I$ the effect is significant in 5\% level while in sections marked with $II$ it is not.}
\label{Fig:RR_sens2_DR}
\end{figure}

We then continued to the analysis under \Dmono for both disease subtypes. The results for the $\sface_{D}$ and $\theta_{D}$ using the DR estimator were given in Figures 2 and 3 of the main text. Figures \ref{Fig:diff_sens2_stand} and \ref{Fig:diff_sens2_iptw} give the corresponding results for standardization and IPTW methods. Figures \ref{Fig:RR_sens2_stand}, \ref{Fig:RR_sens2_iptw} and \ref{Fig:RR_sens2_DR} give the results for the $\sface_{RR}$ and $\theta_{RR}$ using  standardization, IPTW and DR estimators, respectively. While the range for $\lambda_2$ was [0, 1], we only considered $\lambda_2<0.5$. For larger values the  $\sface_{RR}^{(2)}$ estimators and hence $\theta_{RR}$ estimators became unreasonably large.

\begin{figure}[H]
\centering
\includegraphics[scale=0.4]{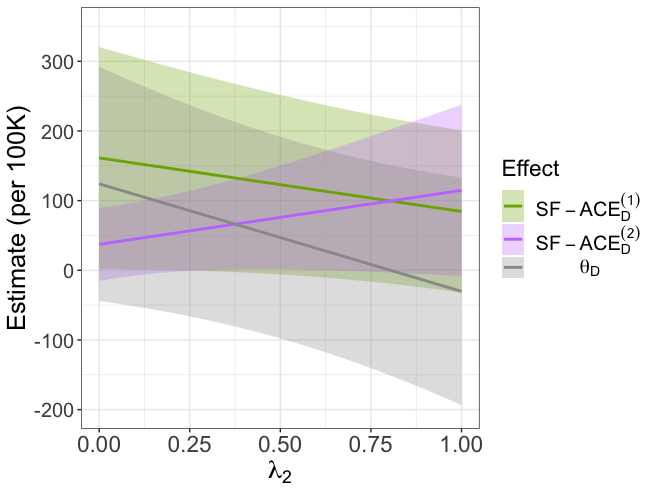}
\captionsetup{font=small}
\caption{Sensitivity analysis for $\sface_D^{(1)}$, $\sface_D^{(2)}$ and $\theta_{D}$ under \Dmono for MSI-high (subtype 2) and \Smono for non-MSI-high  (subtype 1) CRC subtypes. The figure presents DR estimates as a function of $\lambda_2= \Pr[Y^{(1)}(1) = 1|Y^{(2)}(0) = 1]$. The shadows represent 95\% Wald-type confidence intervals connected continuously for clarity of presentation.}
\label{Fig:diff_sens_DR_lambda_2}
\end{figure}

\begin{figure}[H]
\includegraphics[width=\textwidth,height=\textheight,keepaspectratio]{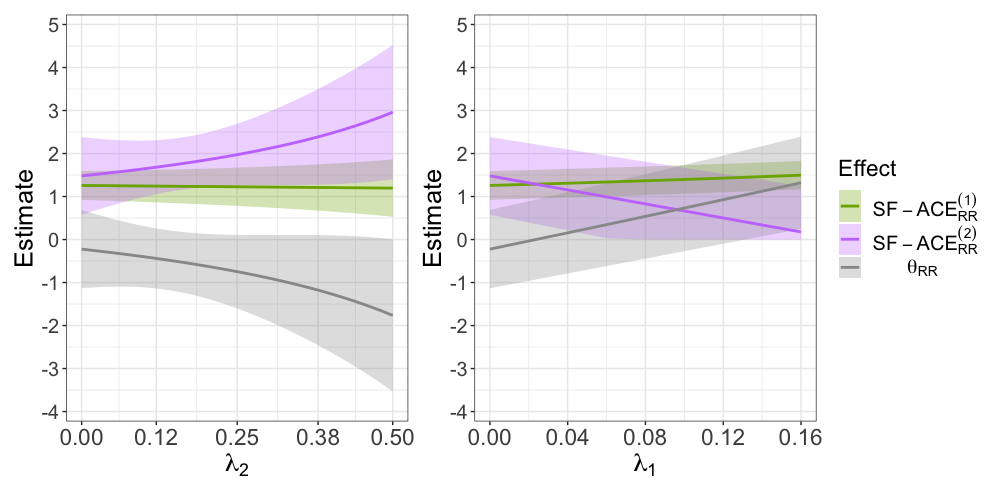}
\captionsetup{font=small}
\caption{Sensitivity analysis for $\sface_{RR}^{(1)}$, $\sface_{RR}^{(2)}$ and $\theta_{RR}$ under \Dmono for MSI-high (subtype 2) and \Smono for non-MSI-high  (subtype 1) CRC subtypes on the left panel, and under \Smono for MSI-high (subtype 2) and \Dmono for non-MSI-high  (subtype 1) CRC subtypes on the right panel. The figure presents DR estimates as a function of $\lambda_2= \Pr[Y^{(1)}(1) = 1|Y^{(2)}(0) = 1]$ on the left panel and $\lambda_1= \Pr[Y^{(2)}(1) = 1|Y^{(1)}(0) = 1]$ on the right panel. The shadows represent 95\% Wald-type confidence intervals connected continuously for clarity of presentation.}
\label{Fig:RR_sens_DR}
\end{figure}

Additionally, we give the results under \Dmono for one subtype and \Smono for the other. The results for the $\sface_{D}$ and $\theta_{D}$ under the more plausible scenario of \Dmono for subtype 1 and \Smono for subtype 2 are given in Figure 2 of the main text. For completeness, Figure \ref{Fig:diff_sens_DR_lambda_2} gives the results for the $\sface_{D}$ and $\theta_{D}$ under \Dmono for subtype 2 and \Smono for subtype 1. Figure \ref{Fig:RR_sens_DR} gives the results for $\sface_{RR}$ and $\theta_{RR}$. The $\sface^{(1)}_{RR}$ estimator was relatively stable while $\sface^{(2)}_{RR}$ increased as a function of $\lambda_2$ and decreased as a function of $\lambda_1$.

Finally, we present the results of an analysis after removing participants for which the last available data were before age 70, with and without inverse probability weighting to take them into account (Tables \ref{Tab:DataResKeepAllNoWeights} and \ref{Tab:DataResKeepAllWeights})

\begin{table}[H]
\centering
\begin{tabular}{lllllll}
Subtype & Effect & Method & Estimate & SE & 95\%CI \\ 
  \hline
 non-MSI-high & $\sface_{RR}$ & Stand & 1.19 & 0.15  & [0.91, 1.48] \\ 
     & & IPTW & 1.21 & 0.14 &  [0.93, 1.48] \\ 
     & & DR & 1.26 & 0.14  & [0.98, 1.54] \\ 
     & $TE_{RR}$ & Stand & 1.19 & 0.15  & [0.91, 1.48] \\ 
      MSI-high &$\sface_{RR}$ & Stand & 1.39 & 0.49  & [0.43, 2.35] \\ 
    & & IPTW & 1.39 & 0.40 & [0.60, 2.27] \\ 
    & & DR & 1.48 & 0.43  & [0.64, 2.32] \\ 
   & $TE_{RR}$ & Stand & 1.39 & 0.54 & [0.32, 2.46] \\ 
\end{tabular}
\caption{Estimated $\sface_{RR}$  and $TE_{RR}$ of smoking  on the two CRC subtypes under \Smono for both subtypes. Stand: standardization; SE: estimated standard error; 95\%CI: 95\% Wald-type confidence intervals.}
\label{Tab:data_res_Smono_RR}
\end{table}

\begin{table}[H]
\centering
\begin{tabular}{lllllll}
 Scale & Effect & Method & Estimate & SE & 95\%CI \\ 
  \hline
   Diff & $\sface_{D}$ & Stand & 97.3 &  89.6 & [-78.3, 273.2] \\ 
    &  & IPTW & 106.8 & 86.8 &  [-63.7, 276.1] \\ 
    &  & DR & 124.1 &   85.8 & [-44.0, 292.6] \\ 
    &  $TE_{D}$ & Stand & 97.4 & 88.6  & [-76.2, 271.1] \\ 
    RR & $\sface_{RR}$ & Stand & -0.19 & 0.44 & [-1.05, 0.67] \\ 
    &  & IPTW & -0.18 & 0.47 &  [-1.08  0.72] \\ 
    &  & DR & -0.22  &  0.46  & [-1.12, 0.68] \\ 
    &  $TE_{RR}$ & Stand & -0.19  & 0.44  & [-1.05, 0.67] \\ 
\end{tabular}
\caption{Estimated $\theta_D$ and $\theta_{RR}$ of the difference between smoking effects on the two CRC subtypes under \Smono for both subtypes. Stand: standardization; SE: estimated standard error; 95\%CI: 95\% Wald-type confidence intervals.}
\label{Tab:data_res_theta_Smono_RR}
\end{table}

\begin{table}[H]
\centering
\begin{tabular}{llllrrr}
  \hline
  Scale & Subtype & Effect & Method & Estimate & SE & 95\%CI \\ 
  \hline
Diff & non-MSI-high & $\sface_{D}^{(1)}$ & Stand  & 146.68 & 100.47 & [-50.24, 343.60] \\ 
 &&& IPTW & 161.92 & 91.71 & [-17.82, 341.66] \\ 
  &&& DR & 186.87 & 105.60 & [-20.10, 393.83] \\ 
   && $TE_{D}^{(1)}$ & Stand &  146.48 & 91.81 & [-33.48, 326.43] \\ 
  & MSI-high &  $\sface_{D}^{(2)}$ & Stand & 37.98 & 28.58 & [-18.03, 93.99] \\ 
 &&& IPTW & 40.97 & 36.03 & [-29.64, 111.58] \\ 
   &&& DR & 44.01 & 30.45 & [-15.67, 103.69] \\ 
   && $TE_{D}^{(2)}$ & Stand & 37.62 & 30.49 & [-22.15, 97.38] \\ 
 RR & non-MSI-high &  $\sface_{RR}^{(1)}$ & Stand & 1.18 & 0.15 & [0.89, 1.47] \\ 
  &&& IPTW & 1.21 & 0.15 & [0.92, 1.49] \\ 
  &&& DR &  1.25 & 0.13 & [0.99, 1.51] \\ 
  && $TE_{RR}^{(1)}$ & Stand & 1.18 & 0.15 & [0.88, 1.48] \\ 
   & MSI-high & $\sface_{RR}^{(2)}$& Stand & 1.38 & 0.39 & [0.60, 2.15] \\ 
 &&& IPTW & 1.43 & 0.42 & [0.61, 2.25] \\ 
  &&& DR & 1.48 & 0.58 & [0.35, 2.61] \\ 
  && $TE_{RR}^{(2)}$ & Stand &  1.38 & 0.43 & [0.54, 2.22] \\ 
   \hline
\end{tabular}
\caption{Estimated $\sface$ and $TE$ of smoking on the two CRC subtypes under \Smono for both subtypes, while removing participants for which the last available data were before age 70. Stand: Standardization; $\hat{SE}$: estimated standard error; 95\%CI: 95\% confidence interval.}
\label{Tab:DataResKeepAllNoWeights}
\end{table}

\begin{table}[H]
\centering
\begin{tabular}{llllrrr}
  \hline
  Scale & Subtype & Effect & Method & Estimate & SE & 95\%CI \\ 
  \hline
Diff & non-MSI-high & $\sface_{D}^{(1)}$ & Stand & 130.66 & 98.68 & [-62.75, 324.07] \\ 
  &&& IPTW & 145.93 & 91.88 & [-34.16, 326.02] \\ 
   &&& DR & 171.47 & 105.58 & [-35.45, 378.40] \\ 
   && $TE_{D}^{(1)}$ & Stand & 130.48 & 94.42 & [-54.58, 315.54] \\ 
   & MSI-high &  $\sface_{D}^{(2)}$ & Stand & 35.73 & 35.73 & [-34.31, 105.77] \\ 
   &&& IPTW & 38.75 & 35.93 & [-31.67, 109.17] \\ 
   &&& DR & 41.67 & 30.97 & [-19.03, 102.37] \\ 
   && $TE_{D}^{(2)}$ & Stand & 35.39 & 36.27 & [-35.70, 106.48] \\ 
  RR & non-MSI-high &  $\sface_{RR}^{(1)}$ & Stand & 1.16 & 0.15 & [0.87, 1.45] \\ 
   &&& IPTW & 1.20 & 0.15 & [0.91, 1.48] \\ 
   && & DR & 1.23 & 0.13 & [0.97, 1.48] \\ 
   && $TE_{RR}^{(1)}$ & Stand & 1.16 & 0.15 & [0.87, 1.45] \\ 
   & MSI-high & $\sface_{RR}^{(2)}$& Stand & 1.35 & 0.45 & [0.47, 2.23] \\ 
   &&& IPTW & 1.42 & 0.41 & [0.61, 2.23] \\ 
   &&& DR & 1.45 & 0.56 & [0.34, 2.55] \\ 
   && $TE_{RR}^{(2)}$ & Stand & 1.35 & 0.55 & [0.27, 2.43] \\ 
   \hline
\end{tabular}
\caption{Estimated $\sface$  and $TE$ of smoking on the two CRC subtypes under \Smono for both subtypes, while removing participants for which the last available data were before age 70 and using inverse probability weighting to take them into account. Diff: defined on the difference scale, RR: defined on the RR scale. Stand: Standardization; $\hat{SE}$: estimated standard error; 95\%CI: 95\% confidence interval.}
\label{Tab:DataResKeepAllWeights}
\end{table}

\end{document}